\documentclass[review]{elsarticle}

\usepackage[margin=1in]{geometry}

\usepackage{lineno}
\usepackage{amsmath}
\usepackage{subfig}
\usepackage[toc,page]{appendix}
\usepackage{blindtext}
\usepackage{natbib}

\usepackage{relsize}
\usepackage{latexsym}
\usepackage{graphicx}
\usepackage{siunitx}
\usepackage{booktabs}
\usepackage{comment}
\usepackage{makecell}
\usepackage{relsize}
\usepackage{array,multirow,tabularx}
\usepackage{appendix}
\usepackage{xcolor}

\usepackage[utf8]{inputenc}
\usepackage[T1]{fontenc}
\usepackage{lmodern}
\usepackage[figurename=Fig.,labelfont=bf,labelsep=period]{caption}
\usepackage{newtxtext,newtxmath}
\usepackage[citecolor=black,linkcolor=black]{hyperref}
\usepackage{tablefootnote}

\DeclareMathOperator{\mdiv}{div}
\DeclareMathOperator{\Div}{Div}

\modulolinenumbers[5]

\journal{Journal of the Mechanics and Physics of Solids}

\biboptions{authoryear}

\begin{document}

\begin{frontmatter}

\title{
Physics and chemistry-based constitutive framework for thermo-chemically aged elastomer using phase-field approach}

\author{Aimane Najmeddine$^{1}$}
\author{Maryam Shakiba\corref{mycorrespondingauthor}$^{1}$}
\cortext[mycorrespondingauthor]{Corresponding author.}
\ead{mshakiba@vt.edu}
\address{$^{1}$Department Civil and Environmental Engineering, Virginia Tech, USA}

\begin{abstract}
We propose a physics and chemistry-based constitutive framework to predict the stress responses of thermo-chemically aged elastomers and capture their brittle failure using the phase-field approach. High-temperature aging in the presence of oxygen causes the macromolecular network of elastomers to undergo complex chemical reactions inducing two main mechanisms: chain-scission and crosslinking. Chemical crosslinking contributes to the stiffening behavior characterizing the brittle response of aged elastomers. In this work, we first modify the Helmholtz free energy to incorporate the effect of thermo-chemically-driven crosslinking processes. Then, we equip the constitutive description with phase-field to capture the induced brittle failure via a strain-based criterion for fracture. We show that our proposed framework is self-contained and requires only four main material properties whose evolution due to thermo-chemical aging is characterized entirely by the change of the crosslink density obtained based on chemical characterization experiments. The developed constitutive framework is first solved analytically for the case of uniaxial tension in a homogeneous bar to highlight the interconnection between all four material properties. Then, the framework is numerically implemented within a finite element (FE) context via a user-element subroutine (UEL) in the commercial FE software Abaqus to simulate more complicated geometries and loading states. The framework is finally validated with respect to a set of experimental results available in the literature. The comparison confirms that the proposed constitutive framework can accurately predict the mechanical response of thermo-chemically aged elastomers. Further numerical examples are provided to demonstrate the effects of evolving material properties on the response of specimens containing pre-existing cracks. 
\end{abstract}


\end{frontmatter}



\section{Introduction}

During the last few decades, elastomers have shown to be extremely advantageous in a number of structural applications across multiple industries. During their service life, elastomers undergo a variety of chemical and mechanical changes that degrade their structural capacity. In particular, exposure to thermal loads (i.e., thermo-chemical aging) causes elastomers to degenerate into weaker materials that can break with minimal mechanical impact. Therefore, it is imperative to investigate how elastomers respond to thermal degradation to better meet structural demands.

Exposure of elastomers to elevated temperatures in the presence of oxygen (i.e., thermo-oxidation) directly affects their mechanical properties. Oxygen acts as a catalyst for the chemical aging of elastomers leading to a progressive alteration of their chemical composition through two main competitive mechanisms: chain scission and crosslink formation (e.g., \cite{Blum1951,GILLEN1995149,COLIN2004309,SHAW20052758,COLIN2007886,Budzien2008,petrikova2011influence,spreckels2012investigations,COQUILLAT20071326,Wineman2009}). The relative rate of chain-scission and crosslink formation is essentially what determines whether the material becomes more ductile or more brittle \citep{celina2013review}. The literature agrees that in most elastomers, rubber chains tend to crosslink under thermo-oxidative conditions, leading to an increase in the modulus and the hardening with embrittlement \citep{WISE1995403,WISE19971929,WISE1997565,CELINA1998493,hamed1999tensile,Celina2000,CELINA2000171,SHAW20052758,celina2013review,JOHLITZ2014138}.

Since the presence of oxygen plays a crucial role in the degradation of elastomeric materials, it is worth distinguishing between two mechanisms in which oxygen can affect material network: \textit{i}) chemical diffusion for samples that are sufficiently thick, or \textit{ii}) through homogeneous distribution for samples with a thickness of approximately 1~$mm$ or smaller \citep{Blum1951}. In applications where test specimens are sufficiently thin, there is enough oxygen available and its distribution is homogeneous such that the aging process is not limited by diffusion \citep{LION20121227}. In this study, we assume sufficiently thin samples for which oxygen is homogeneously distributed and adopt the term thermo-chemical aging instead of "thermo-oxidation" to distinguish between the two scenarios mentioned above. See \cite{SHAW20052758} and \cite{steinke2011numerical} for examples when specimen thickness is large enough such that oxygen diffusion becomes the limiting factor and its implementation must be considered. 

Several researchers have developed analytical and numerical methods to predict the responses of thermo-chemically aged elastomers considering chemical reactions and mechanical coupling (e.g., \citep{achenbach2003finite,SHAW20052758,pochiraju2006modeling,GIGLIOTTI2011431,steinke2011numerical,johlitz2011chemical,Johlitz2013,JOHLITZ2014138,SHAKIBA20144260,SHAKIBA201653,Lejeunes2018constitutive,KONICA2020103858}). Time-temperature equivalence principles based on the Arrhenius relationship have been extensively used to predict the mechanical, physical, and chemical properties of thermo-chemically aged elastomers (e.g., \cite{WISE1995403,GILLEN200325,hassine2014time}). Moreover, phenomenological and thermodynamic-based frameworks were proposed to combine diffusion and reaction expressions to link the mechanical responses to chemical kinetics (e.g., \cite{WISE19971929,WISE1997565,LION20121227,WinemanShaw2019,KONICA2021104347}). Furthermore, micro-mechanical constitutive equations based on statistical mechanics of polymer structure have been introduced (e.g., \cite{MOHAMMADI2020109108,MOHAMMADI20191,Beurle2020,KONICA2021104347}). Recently, \cite{shakiba2021physicsbased} proposed a self-contained constitutive relationship to predict the stiffening response of thermo-oxidatively aged elastomers based solely on the evolution of the macromolecular network characterized by the change in the crosslink density. A similar approach was employed in the work of \cite{najmeddine2021physics} who proposed a stand-alone constitutive framework to capture the mechanical response of photo-oxidatively aged semi-crystalline polymers based on the change in the polymer's crystallinity and mass loss.


Most of the works listed above lacked the important consideration of predicting failure of elastomers during thermo-chemical aging. To take fracture into account, \cite{dal2009micro} proposed a micro-mechanical model based on a series of Langevin-type springs and a bond potential representing the inter-atomic bond energy acting on the chain. The authors used a micro-sphere description for scale transition and predicted the fracture in oxidized rubbers under biaxial loading. \cite{volokh2007,volokh2010modeling,volokh2017loss} introduced the energy limiter concept to limit the stored energy in aged elastomers when subjected to mechanical loading, and therefore, described the stress drop and the entire stress-strain response. Researchers also proposed an approach based on the intrinsic defect concept to predict the ultimate stresses and strains when thermo-oxidative aging is involved \citep{nait2012j,hassine2014time,sadeg2017large,abdelaziz2019new,rezig2020thermo}. \cite{abdelaziz2019new} (and more recently \cite{kadri2022unified}) used the stress limiter approach (and energy limiter approach) to predict the aging effects on stresses and strains at fracture for rubbers based on changes in molar mass (and concentration of elastically active chains and swelling ratio). However, the authors correlated the involved stiffness and fracture material properties to the evolution of their proposed degradation indicators simply through a fitting procedure. Doing so constrains the constitutive framework and renders it a simple fitting algorithm that is only suitable for the particular scenario upon which fitting was calibrated. Moreover, robust coupling of degradation and mechanical responses using continuum damage mechanics or fracture mechanics approaches is missing. It is therefore imperative to develop efficient, robust, and self-contained constitutive frameworks that can simulate and predict the fracture response of thermo-chemically aged elastomers without the need for fitting parameters.


The phase-field approach, which was first introduced in \cite{francfort1998revisiting}, has attracted increasingly more interest thanks to its capability to simulate complex quasi-brittle material responses. In its mathematical description, the method is based on a variational approach wherein crack initiation and propagation are the direct results of the minimization of an energy potential functional describing the Griffith competition between the bulk elastic energy and the surface fracture energy of the elastomer. An alternative description of the phase-field approach considers the method to fall inherently within the general realm of continuum damage theories wherein damage is measured by a scalar field, i.e., the phase-field, giving rise to a definition of cracks as small zones of high gradients of rigidity and strength, analogous to what is commonly done in continuum damage based formulations. In fact, some researches have argued that the phase-field approach to fracture may be regarded as a legitimate continuum gradient damage theory that can be used to describe crack propagation in elastic solids \citep{spatschek2011phase,duda2015phase}.

Since its first proposal, numerous efforts have been made to model brittle as well as quasi-brittle fracture using the phase-field method \citep{Ambati2015review}. More specifically researchers utilized phase-field to simulate rate-independent crack propagation in rubbery polymers at large strains \citep{MIEHE201493,TALAMINI2018434,MaoAnand18,LI2020193}. \citet{kumar2018fracture,kumar2018config} adopted the phase-field formulation to describe the nucleation and propagation of fracture and healing in elastomers undergoing arbitrarily large quasi-static deformations. Integration of the phase-field approach with multi-physics conditions has also been investigated. \cite{MIEHE2015449} proposed continuum phase-field models for brittle fracture towards fully coupled thermo-mechanical and multi-physics problems at large strains. \citet{KONICA2021104347} employed the theory of transient networks, which was advanced by Vernerey and co-authors \citep{vernerey2017statistically,vernerey2018transient,vernerey2018statistical}, to simulate reaction induced chain-scission and crosslinking and coupled it with phase-field to simulate macroscale damage initiation and propagation in aged polymers under mechanical stress. However, while being extremely advantageous in predicting failure responses of aged polymers, \citet{KONICA2021104347}'s framework contained highly complex mathematical considerations which inevitably gave rise to numerous fitting parameters that lacked any physical meaning.   

The phase-field formulation takes as inputs two main variables: the critical energy release rate describing nucleation of fracture from large pre-existing cracks, and an intrinsic length-scale variable which acts as regularization parameter dictating the width of a smeared crack. Discussion of the physical interpretation of the length-scale has lately been a subject for debate. A few works have sought to leverage the relationship between material strength and the length-scale \citep{pham2011gradient,pham2011issues}. For the simple case of single-deformation states such as uniaxial tension, the phase-field can be regarded as a gradient-damage model where the length-scale becomes a constitutive material property related to the strength of the material (e.g., material tensile strength) \citep{tanne2018crack,marigo2016overview,amor2009regularized}. Moreover, through proper treatment of select features in the phase-field formulation, other fracture criteria can be formulated. One such approach is the strain-based criterion for crack nucleation \citep{MIEHE201493}. A natural byproduct of the strain-based criterion is the establishment of a direct relationship between the length-scale, the strain at fracture, the material stiffness, and the critical energy release rate. The interconnection between the length-scale and the other constitutive material properties makes it possible to examine its evolution during thermo-chemical aging as an intrinsic material property. 

In this work, we aim to describe the mechanical responses of thermo-chemically aged elastomers and predict their brittle failure using a thermodynamically consistent framework coupled with the phase-field approach to fracture. This is achieved by recognizing that thermo-chemical aging affects the response of elastomers in the following manner. First, when an elastomer is subjected to thermo-chemical processes, crosslinking is activated and the elastomer becomes more brittle. Embrittlement induces microscopic cracks on the surface of elastomers leading to their brittle fracture when stretched to a certain level of mechanical deformation. The induced micro-cracks propagate within the elastomer by virtue of a competition between two mechanical quantities: \textit{i}) the elastic energy stored in the bulk; and \textit{ii}) the surface energy dissipated through fracture and the creation of new surfaces. We examine these quantities in detail and incorporate the effects of thermo-chemical aging on their evolution -- which is manifested in the change of the crosslink density -- as follows
\begin{itemize}
    \item The effect on the energy stored in the bulk is captured through proper modification of rubber stiffness as a function of crosslink density in the large-strain network-based constitutive theory describing hyperelastic materials.
    \item The effect on the surface energy dissipated through fracture is captured through modification of the well-known Lake-Thomas derivation of the critical energy release rate as a function of evolving crosslink density.
    \item Crack initiation is captured by establishing a strain-based criterion for fracture. The strain at fracture is expressed in terms of the crosslink density and is therefore considered to be known a priori.
    \item The length-scale is considered as an intrinsic material property and is determined by solving the analytical one-dimensional (1D) formulation of the strong forms. The resulting algebraic equation to be solved will be written in terms of the material stiffness, the critical energy release rate, and the strain at fracture (which are all given in terms of the crosslink density).
\end{itemize}
Hence, the developed framework connects the evolution of the material properties in the constitutive equations to the physio-chemical changes in the rubber network. This connection eliminates the need to conduct mechanical testing on aged elastomers and bypasses the need for extra fitting parameters. This work therefore constitutes a clear contribution to the missing relationship between the macromolecular changes and the mechanical and fracture responses of thermo-chemically aged elastomers.

This manuscript is organized as follows. Section~\ref{sec: preliminaries} summarizes the mathematical notations ascribed to kinematic quantities and establishes the fundamental formulation of the problem upon which subsequent derivations are based. Section~\ref{sec: cont model development} describes the developed constitutive framework incorporating the effects of thermo-chemical aging on the coupled hyperelastic-phase-field response of elastomers. The solution of the developed framework for the case of homogeneous one-dimensional bar under uniform tension is explained in Section~\ref{sec: 1D bar}. Validation versus experimental data from the literature are presented in Section~\ref{sec: validation}. Then in Section~\ref{sec: parametric studies}, we discuss our results and present a few parametric studies on a single notch sample aged for varying aging times. Finally, Section~\ref{sec: Conclusions} concludes with some important remarks and ideas for subsequent future investigations. 



\section{Preliminaries} \label{sec: preliminaries}
Tensorial notation is used in this work. Bold letters indicate a vector or a tensor. The inner product is represented by "$\cdot$" and for any two tensors, $\mathbf{A} \cdot \mathbf{B}$, the summation is over the components of the right tensor (e.g., the inner product of two second-order tensors is $\mathbf{A} \cdot \mathbf{B} =tr(\mathbf{A}^T \mathbf{B})$, and for any two vectors, the product is $\mathbf{a} \cdot  \mathbf{b}=\mathbf{a}\mathbf{b}^T$ where the superscript $^T$ indicates tensor or vector transpose). The time rate of change of a quantity in the material configuration (i.e., Lagrangian configuration) is known as a material derivative ($D/{Dt}$) and is indicated by a superimposed dot, whereas the time rate of change of a quantity in the spatial configuration (i.e., Eulerian configuration) is known as the spatial derivative ($\partial/\partial{t}$) and is indicated by a prime sign. Additionally, $\Div$ and $\mdiv$ represent the material and spatial divergence operators, respectively. Finally, ${\nabla _\mathbf{X}}(.)=\frac{{\partial (.)}}{{\partial \mathbf{X}}}$ and ${\nabla _\mathbf{x}}(.)={\nabla}(.)=\frac{{\partial (.)}}{{\partial \mathbf{x}}}$ are the material and spatial gradient operators, respectively.

The problem solved in this work is formulated as follows. Consider an elastomeric body $\Omega_0$ identified with the region of space it occupies within a fixed reference configuration. Denote by $\mathbf{X}$ the location of an arbitrary point in $\Omega_0$ and by $\Gamma_0$ the boundary region of the body with the outward unit normal vector denoted by $\mathbf{m}$. A smooth one-to-one motion mapping can subsequently be defined as $\mathbf{x} = \chi (\mathbf{X},t)$ giving the position of the point at the current configuration for a given time $t \in \mathbb{R}_+$ representing the temporal location. The deformation gradient can then be determined as $\mathbf{F^s} = \nabla \chi (\mathbf{X},t)$. We also define the displacement field $\mathbf{u(X,t)}$ as the difference of the position vector in the reference configuration from the position vector at the current configuration: $\mathbf{u(X,t)} = \mathbf{x} - \mathbf{X}$. Essential displacement boundary conditions are prescribed on ${\Gamma_0}_u$ whereas natural displacement boundary conditions are prescribed on ${\Gamma_0}_t$ such that ${\Gamma_0}_u \bigcap {\Gamma_0}_t = \varnothing $ and ${\Gamma_0}_u \bigcup {\Gamma_0}_t = \Gamma_0 $. 
Additionally, consider that the body $\Omega_0$ contains a sharp crack $\mathcal{S}$ that is smeared over a localization band $\mathcal{B} \subseteq \Omega_0$ with a corresponding outward unit normal vector $\mathbf{m_{d}}$ in which the damage field (or phase-field) $d(\mathbf{X},t)$ localizes. The damage field takes real values between $[0,1]$ in accordance with classical continuum damage mechanics principles where $d(\mathbf{X},t) = 0$ refers to an intact material with no damage and $d(\mathbf{X},t) = 1$ refers to complete fracture. The corresponding essential and natural boundary conditions are prescribed on ${\Gamma_0}_{du}$ and ${\Gamma_0}_{dt}$, respectively.

\section{Constitutive framework coupling thermo-chemical hyperelasticity and phase-field } \label{sec: cont model development}

In this section, we present a detailed description of the proposed constitutive framework governing the response of thermo-chemically aged elastomers within the context of large deformation solid mechanics coupled with phase-field. Section~\ref{sec: gov diff eq} summarizes the governing differential equations for the problem (i.e., strong form) and highlights the constitutive equations describing hyperelasticity and phase-field. Section~\ref{sec: material properties} presents the proposed approach to incorporate the changes in the macromolecular network due to thermo-chemical aging into the constitutive framework. 

\subsection{Governing differential equations and constitutive description} \label{sec: gov diff eq}

The set of governing partial differential equations to be solved for the solid medium with evolving damage are
\begin{equation} \label{Eq_solid_balance ref config in manuscript}
\Div \left( {\mathbf{P} ^s} \right) + {\rho_0} \left( {\mathbf{f}_0 - {\pmb{\gamma}_0 ^s}} \right)  = 0\,\,{\rm{in}}\,\,\Omega_0 \qquad {\rm{and}} \qquad {{\mathbf{P}^s}^T} \mathbf{m} = {\mathbf{t}_0^s}\,\,{\rm{on}}\,\,{\Gamma_0}_{t}
\end{equation}
\begin{equation} \label{Eq_damage_balance ref config in manuscript}
\Div \left( \mathbf{H} \right) -B = \rho_0 \ddot d \,\,{\rm{in}}\,\,\mathcal{B} \qquad {\rm{and}} \qquad \mathbf{H}  \cdot \mathbf{m_d}= 0 \,\,{\rm{on}}\,\,{\Gamma_0}_{dt}
\end{equation} 
where $\mathbf{P}^s$ is the first Piola-Kirchoff stress tensor of the solid, $\mathbf{f}_0$ is the macroscopic body force vector, ${\pmb{\gamma}_0^s}$ is the acceleration vector, $\mathbf{t}_0^s$ is the macroscopic surface traction, and ${\rho_0}$ is the density of the solid medium. $B$ and $\mathbf{H}$ are the two non-classical quantities representing the internal work of damage (dual to $d$) and the flux vector of internal work of damage (dual to $\nabla d$), respectively  \citep{FREMOND19961083}. It should be mentioned that the higher-order micro-traction at the evolving boundaries of the damaged regions are neglected. The reader is referred to Appendix~\ref{AppendixB virtual power} for a detailed derivation of the governing equations based on the principle of virtual power.

Next, the constitutive equation stating the relationship between the stress and the strain quantities and the one stating the relationship between the internal work of damage and its flux vector to the damage variable must be stipulated. The stress-strain relationship can be given by either one of the many expressions of Helmholtz free energy functionals which describe the large deformation behavior of rubber materials \citep{rivlin1948large,ogden1972large,ArrudaBoyce93,gent1996new,ogden1997non}. The Arruda-Boyce (AB) constitutive description \citep{ArrudaBoyce93} - which will be covered in detail in Section~\ref{sec: hyperelasticity} - will be used in this work. The relationship between the internal work of damage and its flux vector to the damage variable is determined based on a thermodynamic analysis for a solid medium with evolving phase-field. For detailed derivation of the constitutive equations, the reader is referred to Appendix~\ref{AppendixC thermodynamic}. The set of equations to be solved then becomes
\begin{equation} \label{Eq_solid_balance in manuscript}
\Div \left( \mathbf{P}^s \right) + {\rho_0} \left( {\mathbf{f}_0 - {\pmb{\gamma}_0^s}} \right)  = 0\,\,{\rm{in}}\,\,\Omega_0 \qquad {\rm{and}} \qquad {\mathbf{P}^s}^T\mathbf{m} = {\mathbf{t}_0^s}\,\,{\rm{on}}\,\,{\Gamma_0}_t
\end{equation}
\begin{align} \label{Eq_damage_balances in manuscript}
\frac{2l_c G_c}{c_{\alpha}} \Delta d - \rho_0 \omega^{\prime}_{(d)} \frac{\partial \Psi }{\partial \omega } -  \frac{G_c \alpha^{\prime}_{(d)}}{l_c c_{\alpha}}  = \rho_0 \ddot d \,\,{\rm{in}}\,\,\mathcal{B} \quad {\rm{and}} \quad \frac{2l_c G_c}{c_{\alpha}} \nabla d  \cdot \mathbf{m_d} = \mathbf{0} \,\,{\rm{on}}\,\,{\Gamma_0}_{dt}
\end{align}
where $\omega_{(d)}$ and $\alpha_{(d)}$ are two characteristic functions in terms of the phase-field variable denoting the degradation and geometric crack functions, respectively, $G_c$ and $l_c$ are the critical energy release rate and the length-scale, respectively, $\mathbf{P}^s= 2 \partial \Psi_{(\mathbf{C},d)} / \partial \mathbf{C}$ where $\Psi_{(\mathbf{C},d)}=\omega_{(d)} \Psi_{(\mathbf{C})}$ is the damaged Helmholtz free energy, ${\mathbf{C}}$ is the right Cauchy-Green strain tensor, and $c_{\alpha} = 4 \int_0^1 \sqrt{\alpha_{(\beta)}} d\beta$. Note that the Cauchy stress tensor $\mathbf{T}^s$ can be calculated as $J^{s^{-1}}\mathbf{P}^s\mathbf{F}^{s^T}$ where $J^s$ is the determinant of $\mathbf{F}^s$.

Various versions of the phase-field approach exist in the literature depending on the choice of $\omega_{(d)}$ and $\alpha_{(d)}$. The more common version corresponds to the case for which $\omega_{(d)}=(1-d)^2$ and $\alpha_{(d)} = d^2$ \citep{bourdin2000numerical,bourdin2008variational,miehe2010phase,ambrosio1990approximation}. In this version, damage begins to evolve at the onset of load application, disallowing the material to develop within the elastic stage. Variations of $\omega_{(d)}$ and $\alpha_{(d)}$ have since been proposed with the aim of introducing further applicability of the model. An alternative version of the phase-field -- which shall be used in this work -- corresponds to the case for which $\omega_{(d)}$ remains unchanged (i.e., $\omega_{(d)}=(1-d)^2$) but $\alpha_{(d)}$ is linear instead of being parabolic (i.e., $\alpha_{(d)} = d$) \citep{pham2011gradient}. This version of phase-field allows the material to develop elastically up to a certain critical strain level upon which fracture initiates. This means that damage is not allowed to commence until the material has reached a critical energy state wherein enough load bearing chains have been broken causing nucleation of fracture. Such a response is characteristic to the behavior of common unaged elastomers (e.g., natural rubber (NR), styrene butadiene rubber (SBR), etc.) when loaded under uniaxial tension as they show a purely nonlinear elastic response up until rupture. In fact, under severe chemical aging scenarios which cause embrittlement, even aged elastomers show an almost linear elastic response in uniaxial tension until they reach the critical level where they cannot sustain any more loads and fracture nucleates due to bond breakage. Therefore, the use of phase-field approaches with linear crack geometric functions is well-suited as it allows the definition of strain-based criteria which can be employed for accurate prediction of fracture initiation for simple cases of deformation such as uniaxial tension. 

In the subsequent sections, we present the conjectured forms of the quantities required to solve the system of Eqs.~(\ref{Eq_solid_balance in manuscript}) and (\ref{Eq_damage_balances in manuscript}) for a particular aging state. These quantities are: the AB hyperelastic free energy (which will be shown to depend on two micromechanically-motivated material properties), the critical energy release rate, the strain at fracture, and the length scale. We show that the evolution of all of these physical properties during thermo-chemical aging can be captured simply through evolving crosslink density. 


%
\subsection{Material properties for thermo-chemically aged elastomer} \label{sec: material properties}

In this section, we strive to connect the macromolecular network alterations to the macroscopic properties and provide appropriate evolution functions for the material parameters involved in the constitutive framework during thermo-chemical aging. We begin by discussing the changes occurring in the material bulk hyperelastic energy, then we present our proposed approach to incorporate the evolution of the chain network in the description of the critical energy release rate and the strain at fracture. We also discuss the role that the length-scale variable plays in capturing fracture initiation.  


\subsubsection{Bulk hyperelastic energy} \label{sec: hyperelasticity}

In a previous work by the authors \citep{shakiba2021physicsbased}, it was confirmed that the crosslinking events in an elastomer induced by thermo-chemical aging contribute significantly to the changes manifested in the free energy. The authors adopted the AB hyperelastic free energy and concluded that thermo-chemical aging causes the number of Kuhn monomers per chain in the AB description to decrease. On the other hand, the formation of crosslinks between the newly formed short-chains induces more stiffness as the deformation of short chains in a highly crosslinked material is more difficult. The authors were able to predict changes in the stiffness due to thermo-chemical aging by incorporating the evolution of the crosslink density in the material's constitutive law. In this work, we elect to follow the same principle.

The AB constitutive equation accounts for the non-Gaussian nature of the molecular chain stretch and provides an accurate representation of the large deformation behavior of rubber-like materials under different states of loading. An attractive feature of the AB description (besides being micro-mechanically motivated) is that it only requires two physics-based material properties, i.e., the network chain density (or equivalently the rubber shear modulus), and the number of Kuhn monomers to simulate elastomer behavior under various deformation states (i.e., uniaxial, shear, and biaxial). Assuming a near-incompressible configuration, the AB Helmholtz free energy can be expressed as
\begin{equation} \label{eq: Helmholtz function for Arruda full form} 
\Psi_{\left(\mathbf{C}\right)} = \Psi_{AB} \left(\mathbf{C}\right) = \mu_0 N_0\Bigg[ \frac{\lambda_{\rm{chain}_{(\mathbf{C})}}}{\sqrt{N_0}} \mathcal{L}^{-1}\Big(\frac{\lambda_{\rm{chain}_{(\mathbf{C})}}}{\sqrt{N_0}}\Big) + \rm{ln}\frac{\mathcal{L}^{-1}\Big(\frac{\lambda_{\rm{chain}_{(\mathbf{C})}}}{\sqrt{N_0}}\Big)}{\rm{sinh}(\mathcal{L}^{-1}\Big(\frac{\lambda_{\rm{chain}_{(\mathbf{C})}}}{\sqrt{N_0}}\Big))} \Bigg]
\end{equation}
where $\mu_0=n_0 K_B \Theta$ is the rubber shear modulus; $n_0$, $K_B$, and $\Theta$ are the number of chains per unit volume, the Boltzmann constant, and the absolute temperature; $N_0$ is the number of Kuhn monomers per chain, $\mathcal{L}(\cdot) = \rm{coth}(\cdot) - \frac{1}{(\cdot)}$ is the Langevin function whose inverse $\mathcal{L}^{-1}$ is given by several approximations in the literature and is equal to $\mathcal{L}^{-1}(x) = x \frac{3-x^2}{1-x^2}$ according to the Pade approximation for some $x \in \mathbb{R}$, and $\lambda_{\rm{chain}_{(\mathbf{C})}} = \sqrt{\frac{I_{1_{(\mathbf{C})}}}{3}}$ is the relative macro-stretch written as a function of the first invariant of the right Cauchy-Green strain tensor $I_{1_{(\mathbf{C})}}=tr(\mathbf{F^s}^T\mathbf{F^s})$. The effect of thermo-chemical aging on the Helmholtz free energy can be accounted for by describing appropriate evolution functions for the rubber modulus and the number of Kuhn monomers with respect to the change in crosslink density.

\cite{shakiba2021physicsbased} showed that the evolution of the rubber modulus during thermo-chemical aging can be given by the following micro-mechanically motivated expression 
\begin{align} \label{eq: evolution of mu - eq1} 
\begin{split}
 \mu_{(t_a)} & = n_{0} K_B \Theta + \big(\rho^{cr}_{(t_a)} - \rho^{cr}_0\big)R \Theta \\
        & = \mu_0 + \big(\rho^{cr}_{(t_a)} - \rho^{cr}_0\big)R \Theta
\end{split}
\end{align}
where $\rho^{cr}_0$ and $\rho^{cr}_{(t_a)}$ are the crosslink densities of the unaged material (at aging time $t_a=0$) and the aged material (at some later aging time $t_a$), respectively, and $R$ is the natural gas constant. In deriving Eq.~(\ref{eq: evolution of mu - eq1}), it is considered that the increase in the number of the newly formed crosslinks per volume due to aging directly affects the rubber modulus of the material at the corresponding aging state. Note that in Eq.~(\ref{eq: evolution of mu - eq1}), the term $(\rho^{cr}_{(t_a)} - \rho^{cr}_0)$ gives the change in the crosslink density between the primary network configuration and the newly formed network configuration corresponding to some aging time $t_a$. A stiffness-like component is introduced by multiplying the change in the crosslink density which has units of moles per volume by $R$ and $\Theta$.

Next, the total number of crosslinks per volume times the number of Kuhn segments per chain must remain constant in order to satisfy the conservation of mass principle. Therefore, the number of Kuhn monomers per chain at the current state of aging, $N_{(t_a)}$, can be obtained according to \citep{shakiba2021physicsbased}
\begin{align} \label{eq: Chain conservation} 
N_{(t_a)} \rho^{cr}_{(t_a)} = N_0 \rho^{cr}_0 
\end{align}

As a result of the modifications considered above, the final form of the AB hyperelastic constitutive equation taking into account the effect of thermo-chemical aging can be written as a function of the stretch and the current state of aging time as follows
\begin{equation} \label{eq: final Arruda-Boyce} 
\Psi_{AB} \left(\mathbf{C},t_a\right) = \mu_{(t_a)} N_{(t_a)} \Bigg[ \frac{\lambda_{\rm{chain}_{(\mathbf{C})}}}{\sqrt{N_{(t_a)}}} \mathcal{L}^{-1}\Big(\frac{\lambda_{\rm{chain}_{(\mathbf{C})}}}{\sqrt{N_{(t_a)}}}\Big) + \rm{ln}\frac{\mathcal{L}^{-1}\Big(\frac{\lambda_{\rm{chain}_{(\mathbf{C})}}}{\sqrt{N_{(t_a)}}}\Big)}{\rm{sinh}(\mathcal{L}^{-1}\Big(\frac{\lambda_{\rm{chain}_{(\mathbf{C})}}}{\sqrt{N_{(t_a)}}}\Big))} \Bigg]
\end{equation}
where $\mu_{(t_a)}$ and $N_{(t_a)}$ are given by Eqs.~(\ref{eq: evolution of mu - eq1}) and~(\ref{eq: Chain conservation}), respectively. Note that Eq.~(\ref{eq: final Arruda-Boyce}) can also be thought of as being a function of the crosslink density $\rho^{cr}_{(t_a)}$ since $\mu_{(t_a)}$ and $N_{(t_a)}$ are both implicit functions of $\rho^{cr}_{(t_a)}$.  

\subsubsection{Critical energy release rate}

In this section, we focus on the development of an important parameter that shows in the phase-field formulation: the critical energy release rate $G_c$. We present an approach to predict its evolution due to thermo-chemical aging by incorporating the change in the material crosslink density. 

Treatment of $G_c$ as an intrinsic material property in rubbers dates back to the work of \cite{lake1967strength}. \cite{lake1967strength} calculated the critical energy release rate in terms of the molecular structure of the elastomer. Their calculation was based on the statistical mechanics framework governing rubber elasticity. The theory is based on the dissociation energy of a single bond in a monomer unit in a perfectly uniform network, $U$. A perfect network is defined as a network where all chains contain the same number of monomer units, $N$, and have the same displacement length as the mean end-to-end distance corresponding to a real network. In such a network, the critical energy release rate can be obtained by multiplying the energy required to rupture a chain, $N U$, by the number of chains crossing a unit area,  $\frac{1}{2}\bar r n$, such that
\begin{equation} \label{eq: fracture toughness 1} 
 G_c = \frac{1}{2} \bar r n N  U, \quad \text{where} \quad  \bar r =  \sqrt{\frac{8 N}{3 \pi}} l
\end{equation}
where $\bar r$ is the mean end-to-end distance of an ideal chain containing $N$ monomer units each of length $l$, and $n$ is the number of chains per unit volume. The presented mean end-to-end distance can be calculated from the theory of rubber elasticity assuming Gaussian statistics for the probability density per unit volume of a randomly jointed chain \citep{lake1967strength}.
Substituting $\bar r$ into Eq.~(\ref{eq: fracture toughness 1})$_a$ yields
\begin{equation} \label{eq: fracture toughness 3} 
 G_c = \sqrt{\frac{2}{3\pi}} n l N^{\frac{2}{3}} U
\end{equation}
which is the final form of a micromechanically motivated Griffith-type criterion. 



Note that the expression of $G_c$ derived herein is written as a function of the number of monomer units $N$. In the AB description, this parameter corresponds to the number of Kuhn monomers, which was derived earlier as a function of the crosslink density (Eq.~(\ref{eq: Chain conservation})). Additionally, the number of chains per unit volume can also be conveniently written as a function of the crosslink density using the expression $n\big(\rho^{cr}_{(t_a)}\big) = \rho^{cr}_{(t_a)} \mathcal{N}_A$ where $\mathcal{N}_A$ is Avogadro's number. We eventually arrive at an expression of $G_c$ written entirely in terms of the crosslink density $\rho^{cr}_{(t_a)}$, the bond dissociation energy of a single bond $U$, and the length of a monomer unit $l$, i.e., 
\begin{equation} \label{eq: fracture toughness 2} 
 G_c \big(\rho^{cr}_{(t_a)}\big) = \sqrt{\frac{2}{3\pi}} \mathcal{N}_A l U \rho^{cr}_{(t_a)} \big(N_{(t_a)}\big)\big)^{\frac{2}{3}}
\end{equation}

The expression derived above for $G_c$ can be thought of as an evolution function for the critical energy release rate given in terms of the crosslink density achieved at a certain aging state during thermo-chemical aging.

\subsubsection{Length-scale} \label{sec: length-scale}

Proper and self-contained identification methods for the length-scale $l_c$ are lacking in the literature. More so, to the best of the authors' knowledge, there exists currently no study which aimed to characterize the evolution of the length-scale parameter when the material is subject to chemical changes. The reason is because most existing phase-field modeling efforts do not consider the length-scale to be an intrinsic material property, but rather a mere regularization parameter which must be small enough to accurately describe crack front propagation. In fact, much debate exists in the literature concerning whether the length-scale parameter could be treated as an independent material parameter. In this work, we confirm that for the special case of uniaxial tension, a physical meaning could be imparted on to the length-scale variable, provided that a strain-based criterion for fracture is utilized. This is accomplished through the realization that if the strain at fracture is known a priori (which we later propose can be given as a function of the crosslink density state of the aged material), then Eq.~(\ref{Eq_damage_balances in manuscript}) can be solved for the strain at fracture by setting $d=0$ in Eq.~(\ref{Eq_damage_balances in manuscript}), and the length-scale $l_c$ can be subsequently tuned to determine the value that will recover the known strain at fracture.


\subsection{Strain-based criterion for fracture of thermo-chemically aged elastomers } \label{sec: strain-based criterion}

In this work, we propose to employ a strain-based criterion for fracture nucleation in smooth specimen (i.e., elastomers containing no large pre-cracks). Fracture can nucleate in a number of ways inside elastomers. As demonstrated through many experimental results, macroscopic crack nucleation can result from either one or all of the following fashions: $i)$ nucleation in the bulk, $ii)$ nucleation from large pre-existing cracks, or $iii)$ nucleation from the boundary and small pre-existing cracks \citep{kumar2018fracture,kumar2018config}. Fracture nucleation from large pre-existing cracks is well-captured by standard phase-field formulations; however, because such formulations lack the important consideration of material strength, they fail to describe crack nucleation in the bulk of smooth elastomers. This limitation restricts their use for fracture problems dealing with uncut samples or sample with no pre-existing cracks. Nevertheless, in this work, we show that instead of explicitly considering material strength, we can establish a strain-based criterion for fracture that will allow us to conveniently describe crack nucleation in smooth elastomers. Our approach is motivated by the physical understanding that like all materials, elastomers are never perfect and contain inherent microscopic defects; thus when a smooth elastomer is stretched monotonically, fracture will nucleate at a given critical value of the applied stretch from one or more of these pre-existing defects. 


The version of the phase-field approach adopted in this work allows for an elastic regime up to the onset of crack nucleation. Such a formulation is attractive as it provides the ability to construct criteria with thresholds for fracture initiation and nucleation. Particularly, it allows us to construct an energetic criterion with threshold based on the limiting strain, i.e., the strain at fracture. Therefore, with the strain at fracture known, a strain-based criterion can be formulated. In this work, the strain at fracture is computed as a function of the crosslink density of the aged material. 

The assumption that the strain at fracture can be explicitly formulated as a function of the crosslink density is motivated by the work of \cite{rezig2020thermo} who verified that the strain at fracture can be expressed linearly as a function of the square root of the molar mass between two crosslinks. Since the molar mass between two crosslinks is related to the crosslink density through an inverse proportionality, a relationship between the strain at fracture and the crosslink density can easily be constructed. As such, we propose the following self-contained equation relating the true strain at fracture to the crosslink density at a given aging time:
\begin{align} \label{eq: true failure strain} 
\begin{split}
    \varepsilon^{tr}_{b_{(t_a)}} & = \sqrt{ \Bigg[ \frac{ \frac{1}{\rho^{cr}_{(t_a)}} - \frac{1}{\rho^{cr}_{max}} }{\frac{1}{\rho^{cr}_0} - \frac{1}{\rho^{cr}_{max}}} \Bigg]} \varepsilon^{tr}_{b0} 
\end{split}
\end{align}
where the superscript $^{tr}$ is in reference to the true configuration, $\varepsilon^{tr}_{b0}$ is the true strain at fracture corresponding to the unaged state, and $\rho^{cr}_{max}$ is the maximum crosslink density that the material can achieve (it is equivalent to the crosslink density at some maximum aging time $t_{a_{max}}$, i.e., $\rho^{cr}_{(t_{a_{max}})}$. It follows that the associated engineering strain at fracture can be obtained from Eq.~(\ref{eq: true failure strain}) as:
\begin{align} \label{eq: engineering failure strain} 
\begin{split}
    \varepsilon^{eng}_{b_{(t_a)}} = exp \Bigg(  \sqrt{ \bigg[ \frac{\rho^{cr}_0 (\rho^{cr}_{(t_a)}-\rho^{cr}_{max})}{\rho^{cr} (\rho^{cr}_0 - \rho^{cr}_{max})} \bigg] }\varepsilon^{tr}_{b0} \Bigg) - 1
\end{split}
\end{align}
Note that the stretch at fracture can be obtained from Eq.~(\ref{eq: engineering failure strain}) through the simple relationship $\lambda_{b} {(t_a)} = \varepsilon^{eng}_{b_{(t_a)}} + 1$.

\begin{figure*}[h!bt]
    \centering
        \includegraphics[scale=0.57]{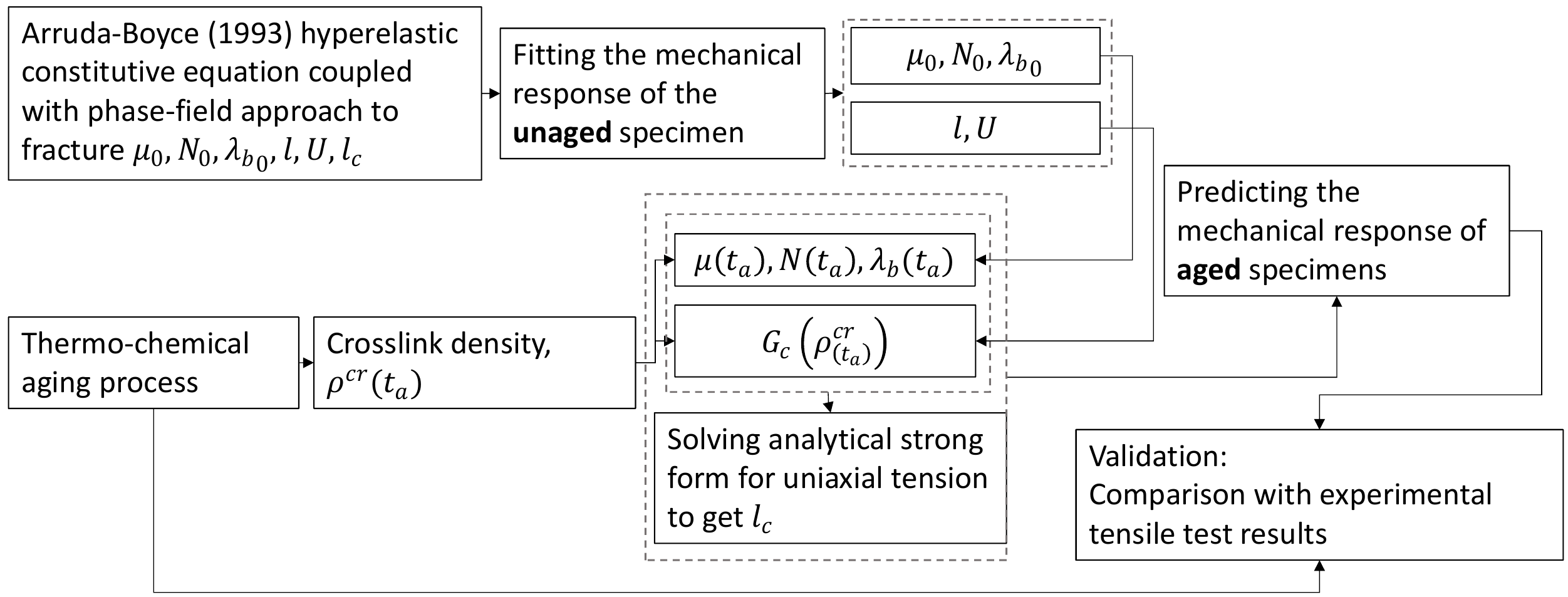}
    \caption{Procedural flowchart for the identification of the material properties and the prediction and validation of the constitutive framework.} 
    \label{fig: flowchart}
\end{figure*}

A summary of the procedure followed to validate the general framework and obtain the material properties involved in this work is illustrated in Figure~\ref{fig: flowchart}. We demonstrate this procedure for the case of homogeneous deformation of a bar under uniform tension in the proceeding section.

\section{Homogeneous case of bar under uniform tension} \label{sec: 1D bar}

In this section, we present the analytical derivation of the proposed framework for the case of a homogeneous bar involving a near-incompressible hyperelastic solid. The bar is assumed to be thermo-chemically aged for varying periods of time and subsequently loaded under uniaxial tension. The present derivation serves to highlight the various steps involved in arriving at the complete stress-strain response of an aged sample from the onset of load application to complete fracture.

Consider an incompressible elastomeric bar subjected to a monotonically increasing tensile stretch. In this case, the deformation gradient is expressed as a function of the applied uniaxial stretch $\lambda$ as follows
\begin{equation}
    \textbf{F}^s(\lambda) = 
    \begin{pmatrix}
        \lambda & 0 & 0\\
        0 & \frac{1}{\sqrt{\lambda}} & 0\\
        0 & 0 & \frac{1}{\sqrt{\lambda}}
    \end{pmatrix}
\end{equation}

The left Cauchy-Green strain tensor can subsequently be written as
\begin{equation}
    \textbf{C}(\lambda) = \textbf{F}^s(\lambda)^T \textbf{F}^s(\lambda) = 
    \begin{pmatrix}
        \lambda^2 & 0 & 0\\
        0 & \frac{1}{\lambda} & 0\\
        0 & 0 & \frac{1}{\lambda}
    \end{pmatrix}
\end{equation}
whose first invariant is given by: $I_{1_{(\lambda)}} =  tr(\textbf{C}) = \frac{2}{\lambda} + \lambda^2$

Eq.~(\ref{eq: final Arruda-Boyce}) can be written in polynomial form using the first five terms of the inverse Langevin function as
\begin{align} \label{eq: example AB} 
\Psi_{AB} (\mathbf{C},t_a) = \mu_{(t_a)} \mathlarger{\sum}_{i=1}^5 c_i  \frac{1}{N_{(t_a)}^{2i-2}} \left( I_{1\mathbf{C}}^i -3^i  \right)
\end{align}
where the constants $c_i$ in Eq.~(\ref{eq: example AB}) are equal to $c_1=\tfrac{1}{2}, c_2= \tfrac{1}{20}, c_3= \tfrac{11}{1050}, c_4= \tfrac{19}{7000}, c_5= \tfrac{519}{673750}$, and $\mu_{(t_a)}$ and $N_{(t_a)}$ are given by Eqs.~(\ref{eq: evolution of mu - eq1}) and (\ref{eq: Chain conservation}), respectively.

In the case of uniaxial tension, Eq.~(\ref{eq: example AB}) can be expressed as a function of the applied stretch as
\begin{align} \label{eq: example AB UT}
\Psi_{AB} (\lambda,t_a) = \mu(t_a) \mathlarger{\sum}_{i=1}^5 c_i  \frac{1}{N(t_a)^{2i-2}} \left( I_1(\lambda)^i -3^i  \right)
\end{align}
Therefore, the first Piola-Kirchhoff stress in uniaxial tension $P^s$ can be computed as
\begin{align} \label{stress}
 P^s(\lambda,t_a) = \frac{\partial\Psi_{AB} (\lambda,t_a)}{\partial\lambda}
\end{align}

For the case of homogeneous damage state, $\Delta d = 0$ (i.e. damage is uniform in the bar) and $\ddot d=0$. Therefore, Eq.~(\ref{Eq_damage_balances in manuscript})a becomes
\begin{align} \label{Eq_damage_balances_example}
\rho_0 \omega^{\prime}_{(d)} \frac{\partial \Psi(\lambda,d) }{\partial \omega } + \frac{G_c {\alpha^{\prime}_{(d)}}}{l_c c_{\alpha}}  = 0  \,\,{\rm{in}}\,\,\mathcal{B}
\end{align}
where $\frac{\partial \Psi(\lambda,d)}{\partial \omega } = \frac{\partial (\omega_{(d)} \Psi_{AB}(\lambda,t_a))}{\partial \omega } = \Psi_{AB}(\lambda,t_a)$ is the undamaged AB Helmholtz free energy. Note that for the version of the phase-field employed in this work, the first derivatives of the degradation and the crack geometric functions with respect to the phase-field variable are given by $\omega^{\prime}_{(d)} = 2(d-1) $ and $\alpha^{\prime}_{(d)} = 1$, respectively, while $c_{\alpha}$ is given by $c_{\alpha}=4 \int_0^1 \sqrt{\alpha_{(\beta)}} d\beta = \frac{8}{3}$. 

Eq.~(\ref{Eq_damage_balances_example}) is the balance equation for the phase-field variable governing the evolution of the damage field inside the body $\mathcal{B}$. The phase-field variable can be solved either analytically or numerically provided that all necessary inputs are known. These inputs are: the crosslink density $\rho^{cr}_{(t_a)}$ for a given aging time $t_a$, the critical energy release rate $G_c \big(\rho^{cr}_{(t_a)}\big)$ corresponding to said crosslink density, and the length-scale $l_c$. The length-scale depends on the material stiffness and its fracture resistance, thus also implicitly on the crosslink density, as shown through Eq.~(\ref{Eq_damage_balances_example}).

The length-scale is determined following the procedure described in section~\ref{sec: length-scale}. To see how this is accomplished, let us simplify Eq.~(\ref{Eq_damage_balances_example}) by substituting the corresponding terms. We obtain
\begin{align} \label{damage_balances_simplified}
 \Psi_{AB}(\lambda_{b},t_a) - \frac{3G_c \big(\rho^{cr}_{(t_a)}\big)}{8l_c} = 0  \,\,{\rm{in}}\,\,\mathcal{B}
\end{align}
where $\lambda_{b}$ is the critical stretch (stretch at failure). Note that the value $d=0$ was substituted for the phase-field variable since fracture will nucleate when $d$ ceases to be identically 0. Thus, for a particular value of the crosslink density, $\Psi_{AB}(\rho,\lambda_{b})$ and $G_c \big(\rho^{cr}_{(t_a)}\big)$ can be determined using Eq.~(\ref{eq: example AB UT}) and Eq.~(\ref{eq: fracture toughness 2}), respectively. Therefore, the length-scale can be tuned such that the resulting value for the critical stretch $\lambda_{b}$ from Eq.~(\ref{damage_balances_simplified}) matches the one obtained through the critical strain in Eq.~(\ref{eq: engineering failure strain}). Figure~\ref{fig: flowchart} illustrates a procedural flowchart for the identification of the material properties and the prediction and validation of the proposed constitutive framework. In the section that follows, we present validations of the proposed constitutive framework applied to the case of an elastomer aged for varying periods of time. 

\section{Validation of the developed constitutive framework} \label{sec: validation}

\subsection{Homogeneous solution}

\cite{rezig2020thermo} conducted a series of experimental studies on the thermo-chemical aging effects in filled SBR. The authors determined the crosslink densities corresponding to various aging times ranging from 0 to 60 days for a series of temperatures. In this paper, we validate the proposed framework versus the case for which aging was performed at $\SI{100} {\celsius}$. 

\begin{figure*}[hbt!]
    \centering
    \subfloat[]{%
        \includegraphics[width=0.4\linewidth]{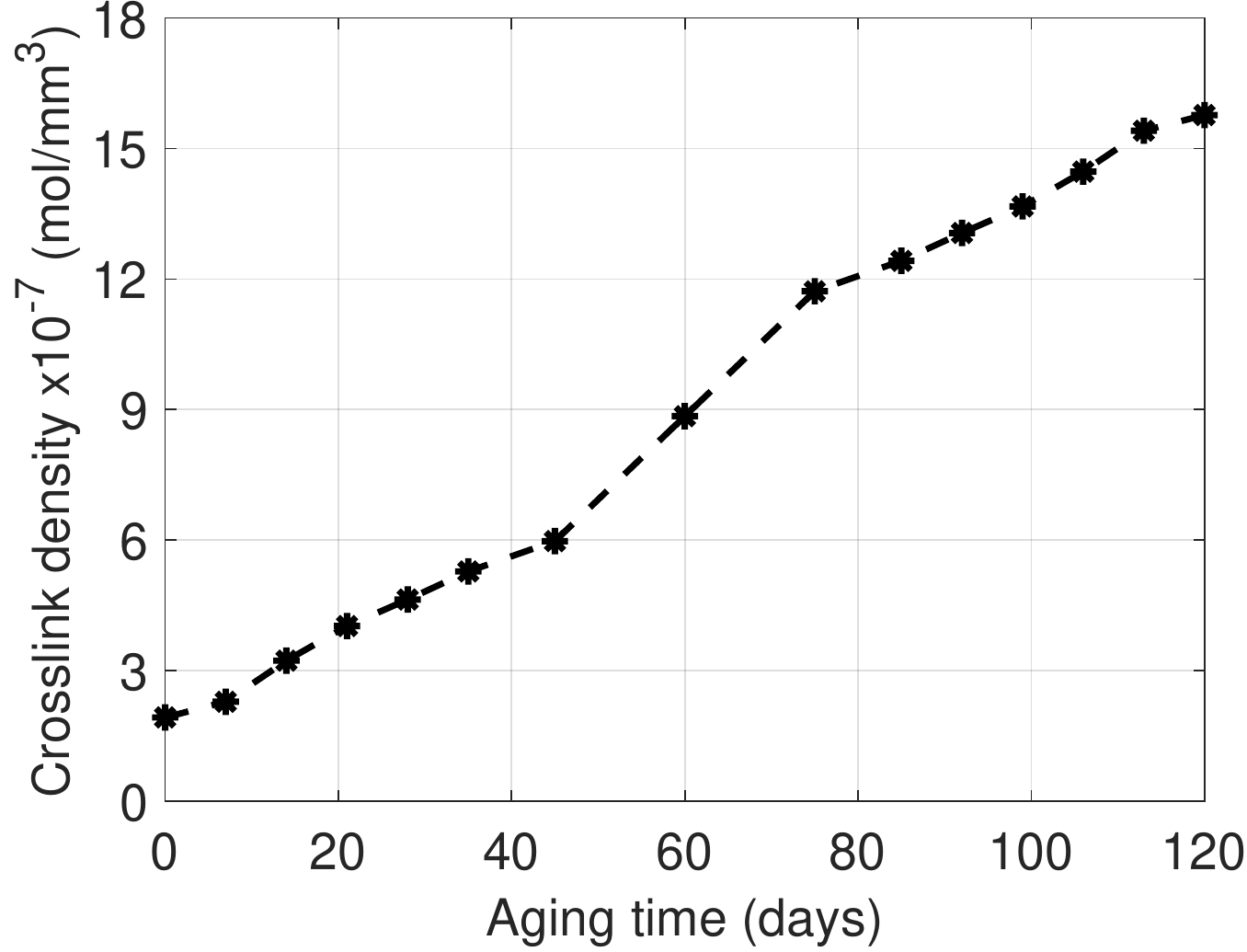}
        \label{fig: crosslink_density}
    }%
    \hspace{0.5cm}
    \subfloat[]{%
        \includegraphics[width=0.4\linewidth]{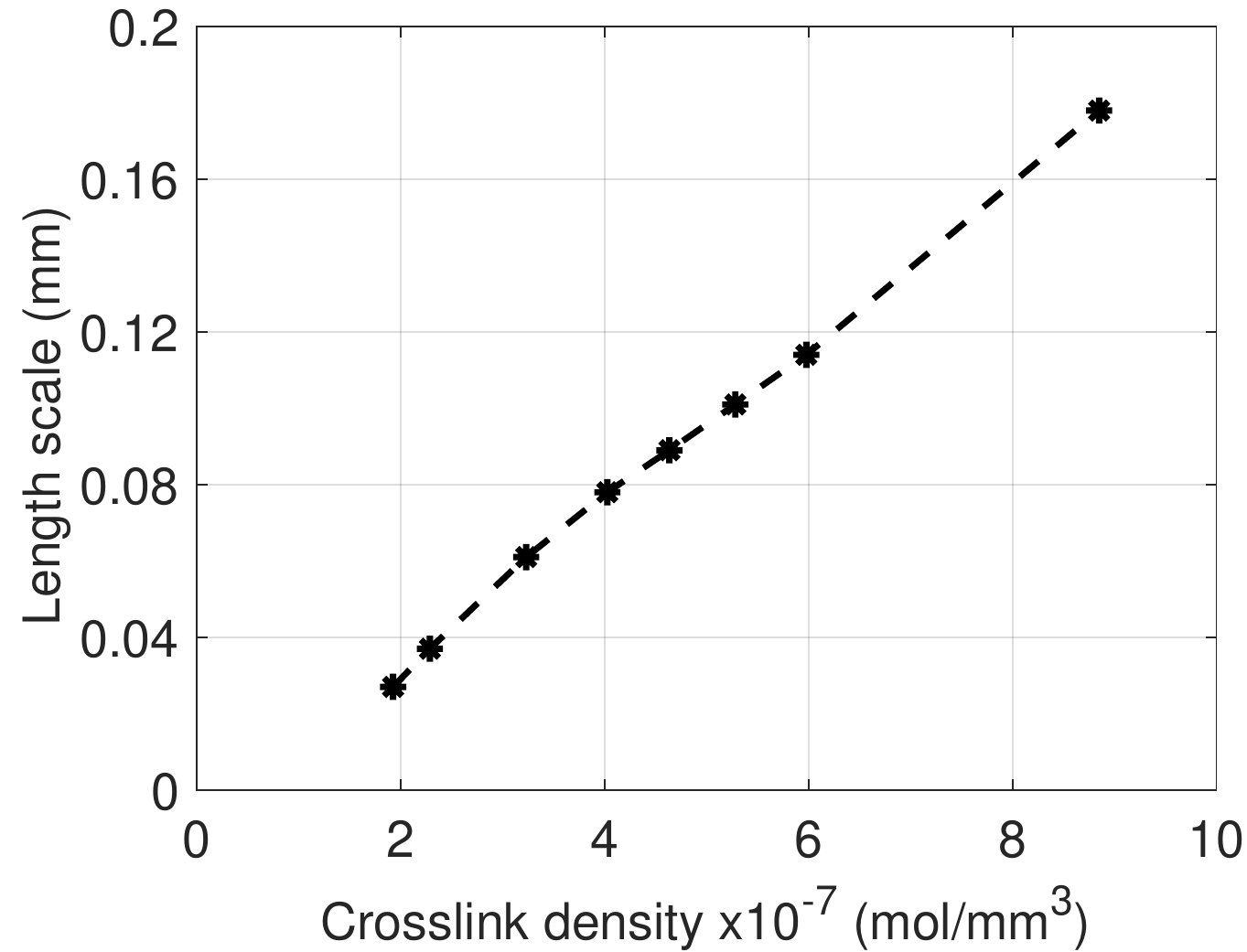}
        \label{fig: length_scale}
    }%
    \caption{a) Evolution of the crosslink density as a function of aging time in an SBR sample thermally aged under $\SI{100} {\celsius}$ \citep{rezig2020thermo}. The values in the [60,120]-day range have been extrapolated linearly based on the available data. b) Evolution of the length-scale $l_c$ as a function of the crosslink density for varying aging times. }
    \label{}
\end{figure*}

First, we need to determine the maximum value of the crosslink density required for Eq.~(\ref{eq: engineering failure strain}). A linear extrapolation procedure was employed and the value corresponding to $t_{a_{max}} = 120$ days, i.e., $ \rho^{cr}_{max} = \rho^{cr}_{(t_{a_{max}})}$ was selected. Figure~\ref{fig: crosslink_density} presents the evolution of the crosslink density as a function of aging time $t_a \in [0,{t_a}_{max}]$. As an example, let us consider the case for which the material is thermally aged for a period of $t_a = 45 $ days under $\SI{100} {\celsius}$. 
Substituting $t_a = 45$ in the expression for $G_c \big(\rho^{cr}_{(t_a)}\big)$ (Eq.~(\ref{eq: fracture toughness 2})) and in the expression for $\Psi_{AB}(\lambda_{b},t_a)$ (Eq.~(\ref{eq: example AB UT})), we can solve for $l_c$ in Eq.~(\ref{damage_balances_simplified}) with $\lambda_{b} = \varepsilon_b^{eng}(t_a) + 1$. For this example, the obtained $l_c$ value was $0.114~mm$. 

The procedure can be extended to the remaining aging times and the $l_c$ corresponding to each aging state can be calculated in a similar fashion. Table~\ref{tb: material properties} summarizes the values for the material properties obtained for the various aging times considered. Note that the dissociation energy $U$ was taken to be the average dissociation energy of the C-C bonds in a monomer unit. For a single monomer, $U$ can be obtained by dividing the molar dissociation energy (which is given in unis of (energy/moles)) by Avogadro’s number. For the problem in hand, a value of $U = 5.779 \times 10^{-9} joule$ was calculated. Additionally, the length of monomer units $l$ in Eq.~(\ref{eq: fracture toughness 2}) was assumed to be constant for all aging times and was therefore calculated based on the response of the unaged configuration. 

The resulting $l_c$ values are plotted as a function of the crosslink density in Figure~\ref{fig: length_scale} for $t_a \in [0,60]$ days. Interestingly, $l_c$ is shown to evolve linearly with respect to the crosslink density. This linear relationship suggests that the length-scale, similarly to the crosslink density, should also evolve in a sigmoidal manner with respect to aging time. This finding is crucial as it sheds light on the evolution of an important parameter in the phase-field characterization of damage in thermo-chemically aged elastomers.

\begin{table}[h!bt]
\centering
\small
\caption{Material properties obtained using the developed framework for the various aging times considered in \cite{rezig2020thermo} for an SBR sample thermally aged under $\SI{100} {\celsius}$.}
\begin{tabular}{ccccc}
        \hline
        \multirow{4}{*}{\thead{Aging time \\ $t_a$ (days)}} & \multicolumn{4}{c}{Material properties} \\ \cmidrule{2-5}
        {} & \multirow{3}{*}{\thead{Rubber modulus \\ $\mu$ (MPa) \\ (Eq.~(\ref{eq: evolution of mu - eq1})) }} & \multirow{3}{*}{\thead{Number of \\ Kuhn monomers $N$ \\ (Eq.~(\ref{eq: Chain conservation}))  }} & \multirow{3}{*}{\thead{Critical energy \\ release rate $G_c$ ($N/mm$) \\ (Eq.~(\ref{eq: fracture toughness 2})) }} &  \multirow{3}{*}{\thead{length-scale \\ $l_c$ ($mm$) \\ (Eq.~(\ref{damage_balances_simplified}))  }} \\ \\ \\
        \hline
        0 & 0.8 & 70 & 9 & 0.027 \\
        7 & 0.91 & 58.9 & 8.27 & 0.037 \\
        14 & 1.21 & 41.7 & 6.96 & 0.061 \\
        21 & 1.45 & 33.5 & 6.23 & 0.078 \\
        28 & 1.64 & 29.1 & 5.81 & 0.089 \\
        35 & 1.84 & 25.5 & 5.44 & 0.101 \\
        45 & 2.06 & 22.6 & 5.11 & 0.114 \\
        60 & 2.95 & 15.2 & 4.20 & 0.178 \\
        \hline
\end{tabular}
\label{tb: material properties}
\end{table}

\begin{figure*}[hbt!]
    \centering
    \subfloat[]{%
            \includegraphics[scale=0.5]{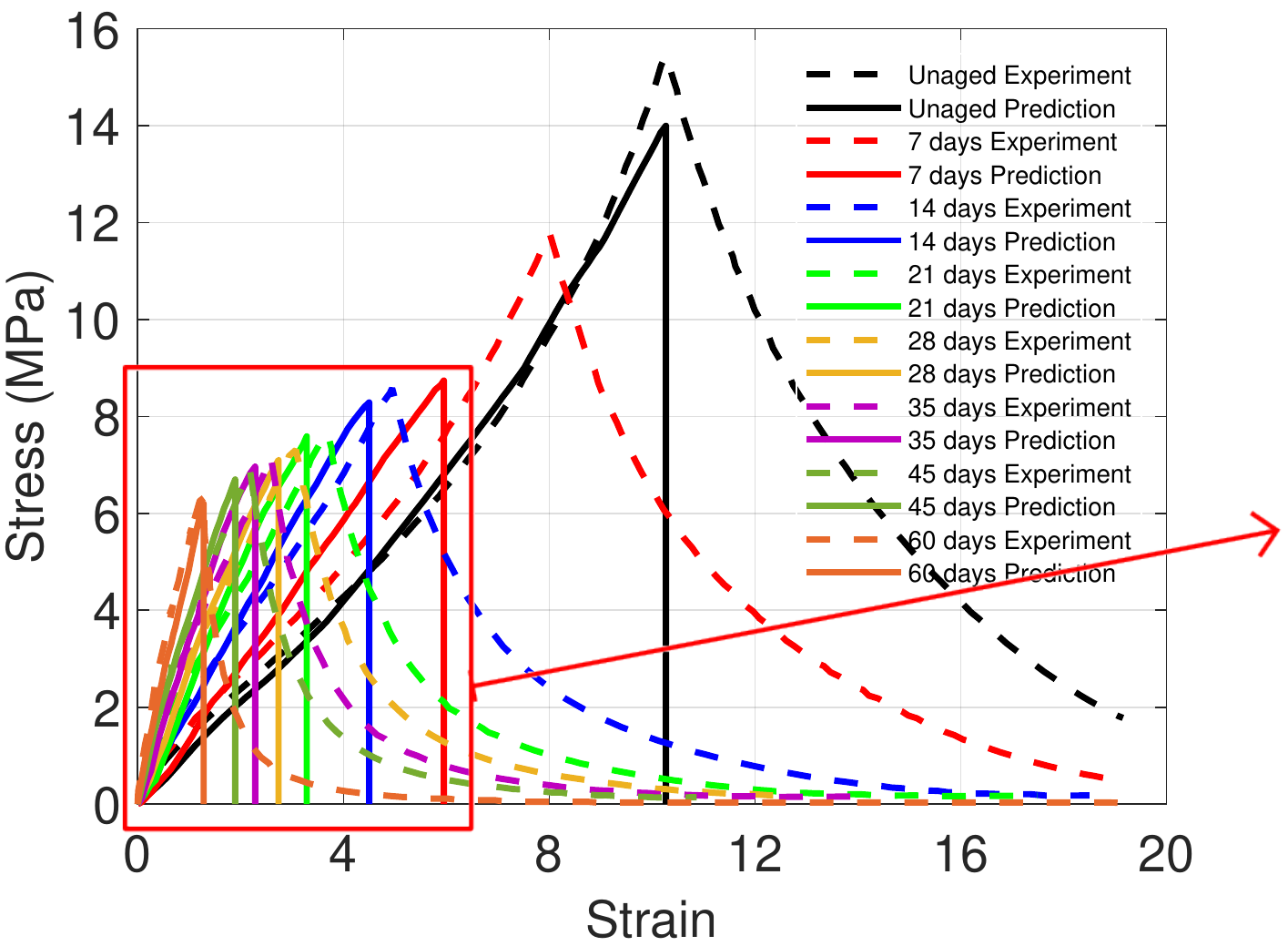}
            \label{fig: full}
    } 
    \hspace{-0.1cm}
    \subfloat[]{%
            \includegraphics[scale=0.5]{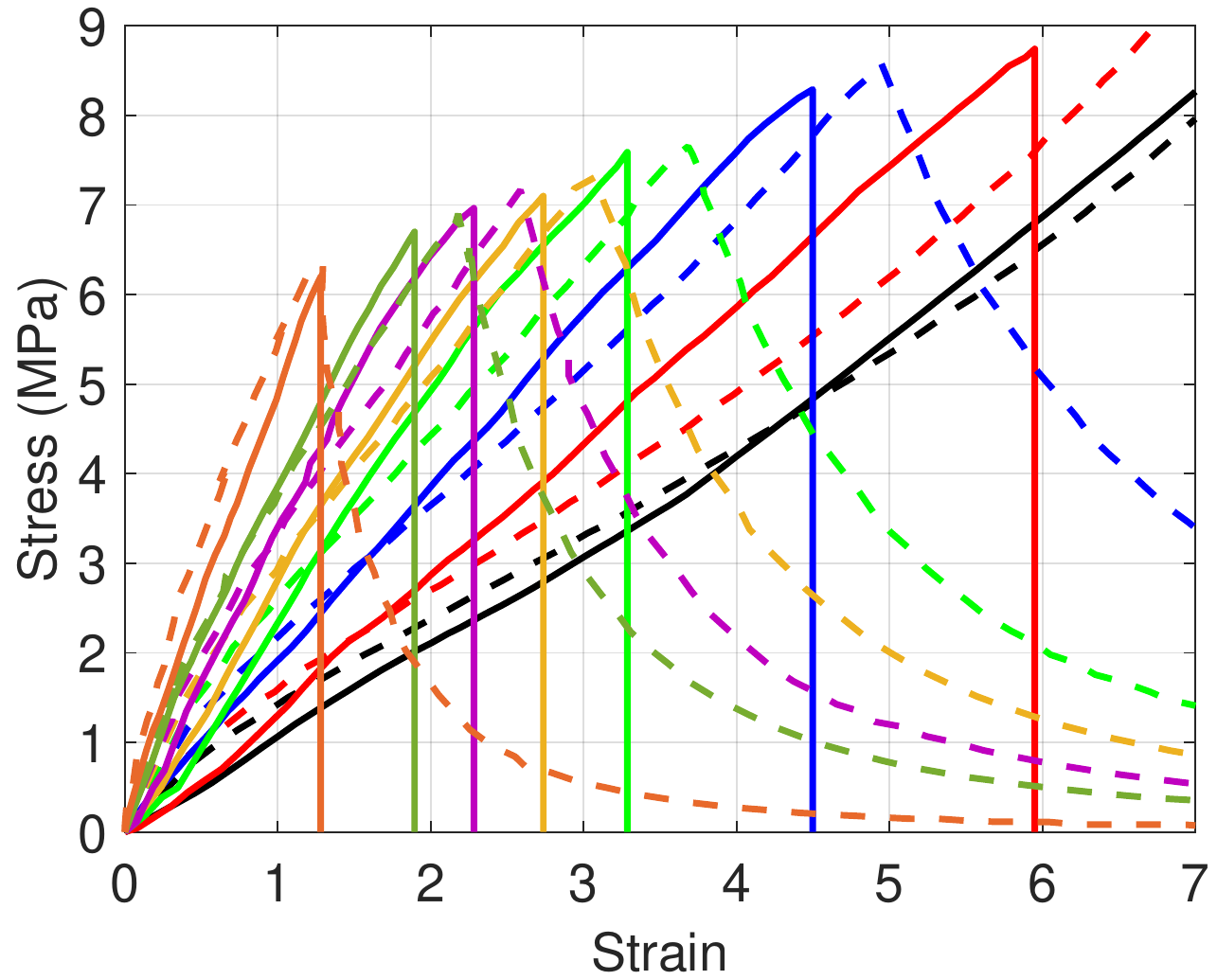}
            \label{fig: zoom}
    }
    \caption{Predictions of the developed constitutive framework using the 1D analytical derivations for the case of uniaxial tension verified against the experimental results for an SBR material aged at $\SI{100} {\celsius}$. Experimental data are reported in \cite{rezig2020thermo}; a) full-range stress-strain responses, b) enlarged-picture of the low-range stress-strain responses }
    \label{fig: prediction_exp_ann}
\end{figure*}

Once the length-scale is determined, the engineering stress-strain response corresponding to each particular aging time can be analytically derived for the homogeneous case of a bar under uniform tension. Figure~\ref{fig: prediction_exp_ann} shows the predictive capability of the developed framework. A very good match between the experimental results and the stress-strain curves calculated using the present approach is achieved. 

\subsection{Finite element solution}

\begin{figure*}[hbt!]
    \centering
        \subfloat[]{%
            \includegraphics[scale=0.6]{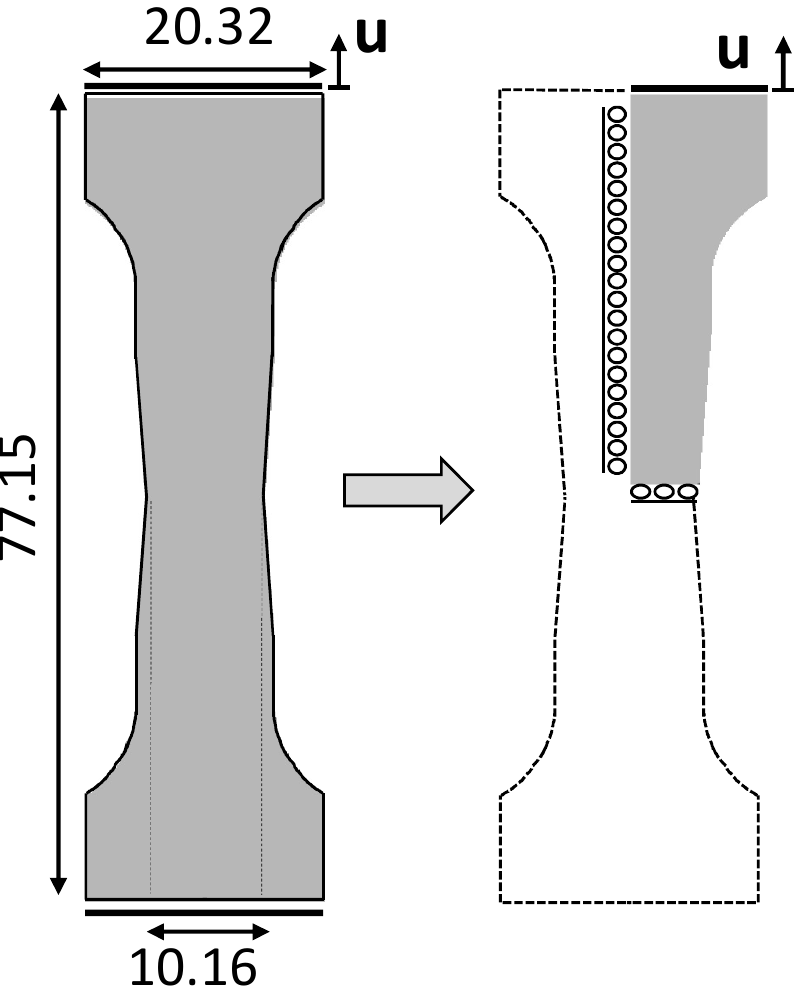} 
            \label{fig: geometry}
        }%
        \hspace{1.5cm}
        \subfloat[]{%
            \includegraphics[scale=0.89]{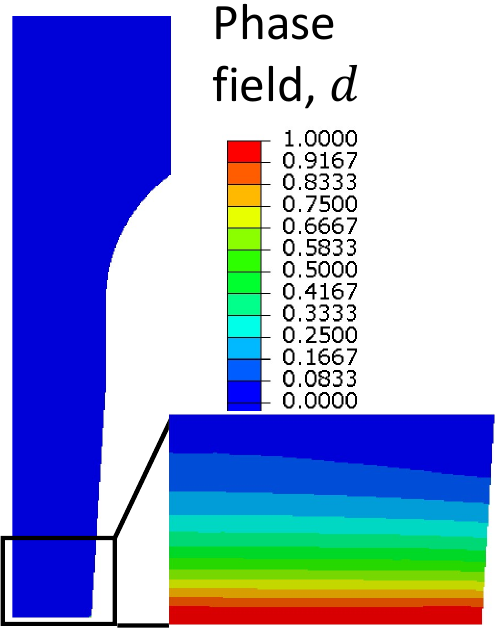} 
            \label{fig: damage_contour}
        }%
    \caption{a) Sample geometry used in the finite element simulations. All dimensions are given in $mm$ unit. For the sake of minimizing the computational cost, only the quarter geometry was used and symmetric boundary conditions were applied on the left and bottom edges; b) contour plot for the phase-field damage variable for an SBR material thermally aged for 45 days at $\SI{100} {\celsius}$. }
\end{figure*}

In this section, we discuss the finite element (FE) solution of the thermo-chemical aging response of a dumbbell-shaped sample axially loaded in tension by a prescribed deformation $u$ (Figure~\ref{fig: geometry}). We focus on the case for which the material was aged for 45 days under $\SI{100} {\celsius}$. Details of the finite element (FE) implementation of the present constitutive framework are attached in Appendix~\ref{Apdx: FE implementation}. The FE simulations were performed on the FE software Abaqus \citep{abaqus2011dassault} via a user-element subroutine (UEL) within a two-dimensional (2D) context. In all simulations, the element size was taken to be $l_c/4$ and plain strain quadrilateral elements were used. The system of governing differential equations was solved using the staggered solution algorithm proposed by \cite{Miehe2010}. To minimize the computational cost associated with the FE simulation, only a quarter of the geometry was used and symmetric boundary conditions were applied on the left and bottom edges as shown in Figure~\ref{fig: geometry}. The material properties for the case $t_a = 45$ days which are presented in Table~\ref{tb: material properties} were used to run the simulations. 

Figure~\ref{fig: damage_contour} illustrates the contour plot for the damage variable highlighting the critical region which experiences extreme damage. Note that the width of the diffuse damage band is governed by the value of $l_c$. Recall that for the present case, i.e., 45 days of aging time, $l_c$ was found to be $0.114~mm$. This is approximately 0.15\% of the specimen dimension. As pointed out in \cite{mandal2019length}, when the length-scale is considered as a material constant (which is the case for the present study) and is small with respect to the dimensions of the sample, both the peak load as well as the damage contour can be very well captured using the phase-field approach adopted here. In this work, we have shown that the obtained $l_c$ values for the varying aging times increases linearly with respect to the crosslink density, and thus in a sigmoidal manner with respect to aging time. It is thus expected that the length-scale would reach a plateau at some maximum aging time. It remains to evaluate whether damage patterns would provide any meaningful conclusions for cases where $l_c$ approaches such limit.

\begin{figure*}[hbt!]
    \centering
        \subfloat[]{%
            \includegraphics[scale=0.5]{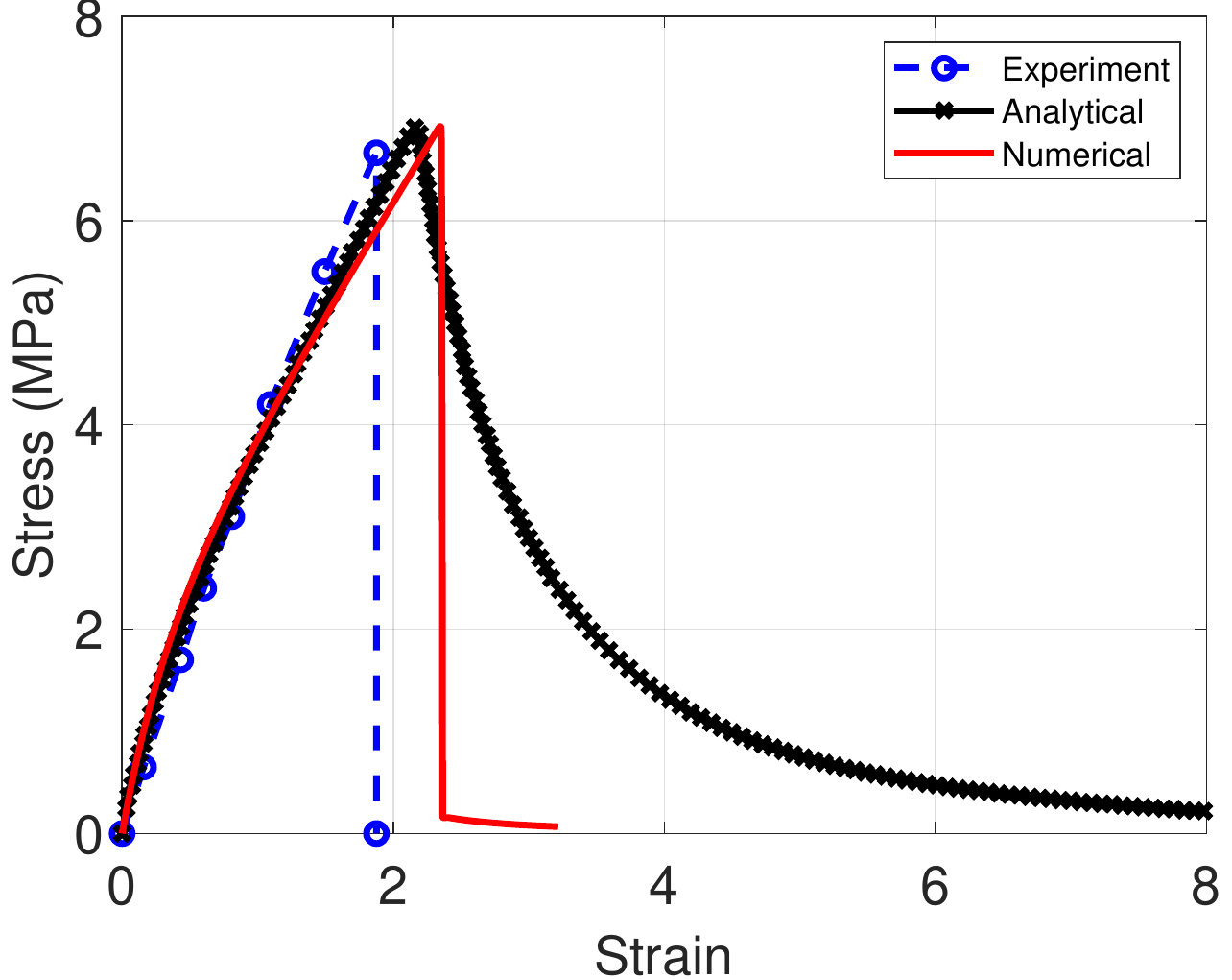} 
            \label{fig: stress_strain_45}
        }%
        \hspace{0.0cm}
        \subfloat[]{%
            \includegraphics[scale=0.5]{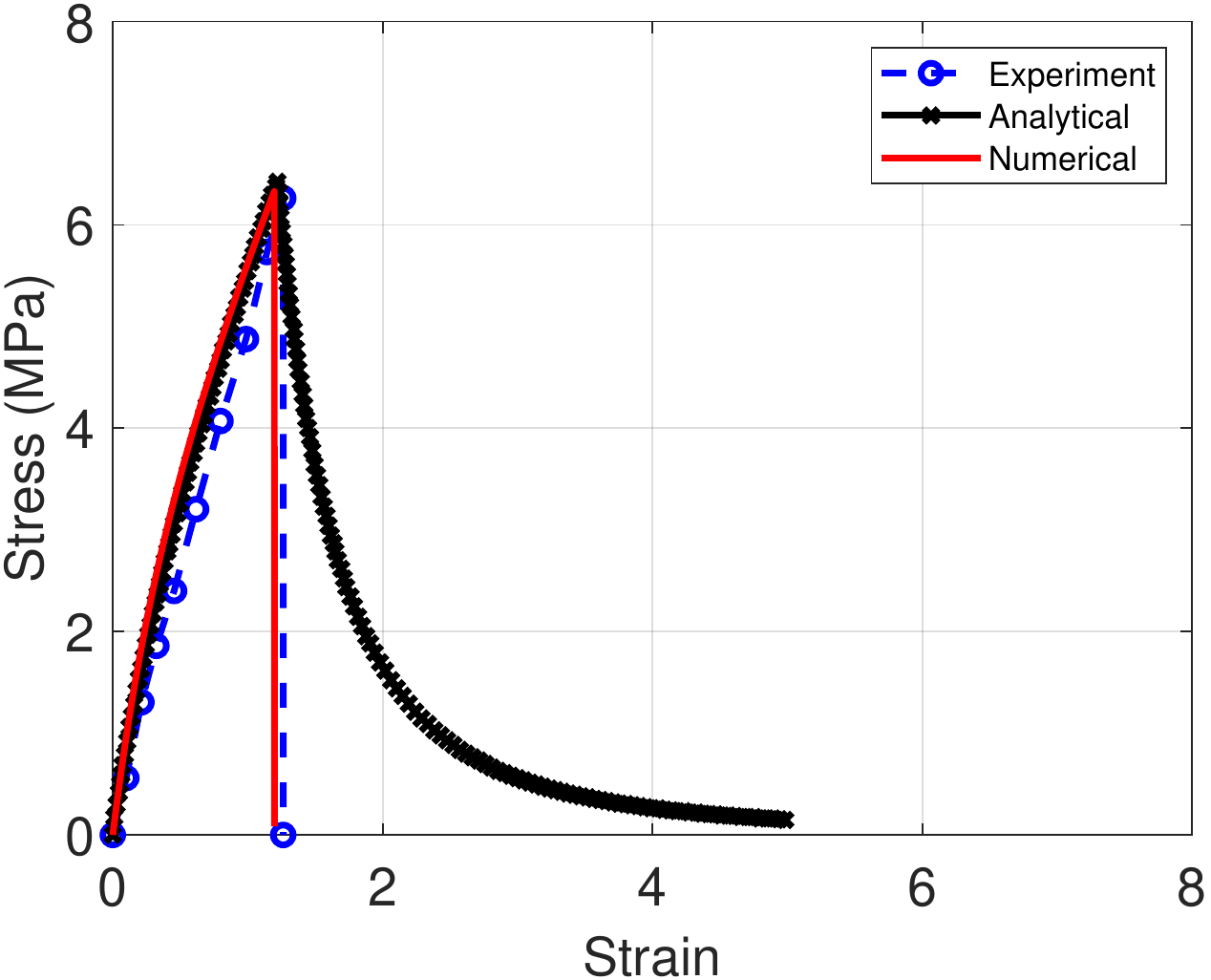} 
            \label{fig: stress_strain_60}
        }%
    \caption{Comparison between the stress-strain responses using the present framework (obtained both analytically and numerically) and the experimental stress-strain curve for an SBR thermally aged for a) 45 days and b) 60 days at $\SI{100} {\celsius}$ \citep{rezig2020thermo}.}
    \label{fig: stress_strain_dumbbell}
\end{figure*}

Figure~\ref{fig: stress_strain_45} demonstrates the comparison between the stress-strain responses using the present framework (obtained both analytically and numerically) and the corresponding experimental response. Additionally, to solidify our argument, we also present in Figure~\ref{fig: stress_strain_60} a similar comparison for the case when $t_a = 60$ days. It can be seen that the framework can predict the responses of both aging times with very high accuracy. Particularly, the increased stiffness due to thermo-chemical aging, the peak stress reached within the aged material, and the strain at fracture linked to the change in the crosslink density are all shown to match highly accurately with the experimental response for both aging states. In fact, treating the length-scale as intrinsic material property and relating the changes in the critical energy release rate and the strain at fracture to the evolution of the crosslink density has proved vastly efficient in capturing the full stress-strain response of the thermo-chemically aged elastomers. The highly predictive capability of the proposed constitutive framework makes the present effort especially attractive as it combines our understanding of how thermo-chemical aging affects the macromolecular structure of the network and the adaptability of phase-field approach to simulate brittle fracture.

\section{Discussion and parametric studies} \label{sec: parametric studies}

\begin{figure*}[hbt!]
    \centering
        \subfloat[]{%
            \includegraphics[scale=1.1]{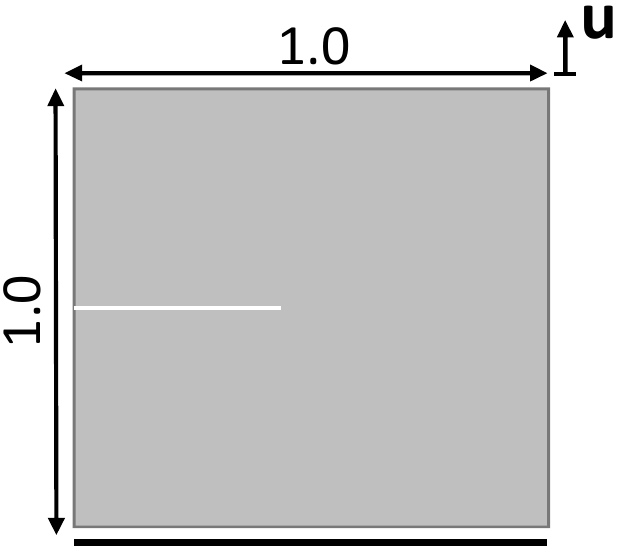} 
            \label{fig: BCSNUT}
        } 
        \hspace{-0.0cm}
        \subfloat[]{%
            \includegraphics[scale=1.1]{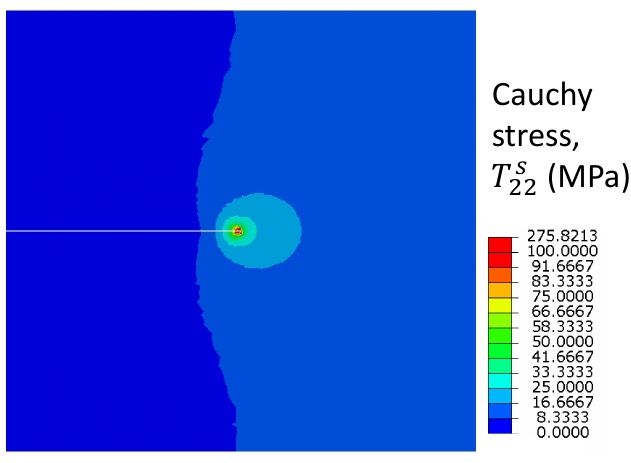} 
            \label{fig: SN_UT_60_S22}
        }%
        \\
        \subfloat[]{%
            \includegraphics[scale=0.75]{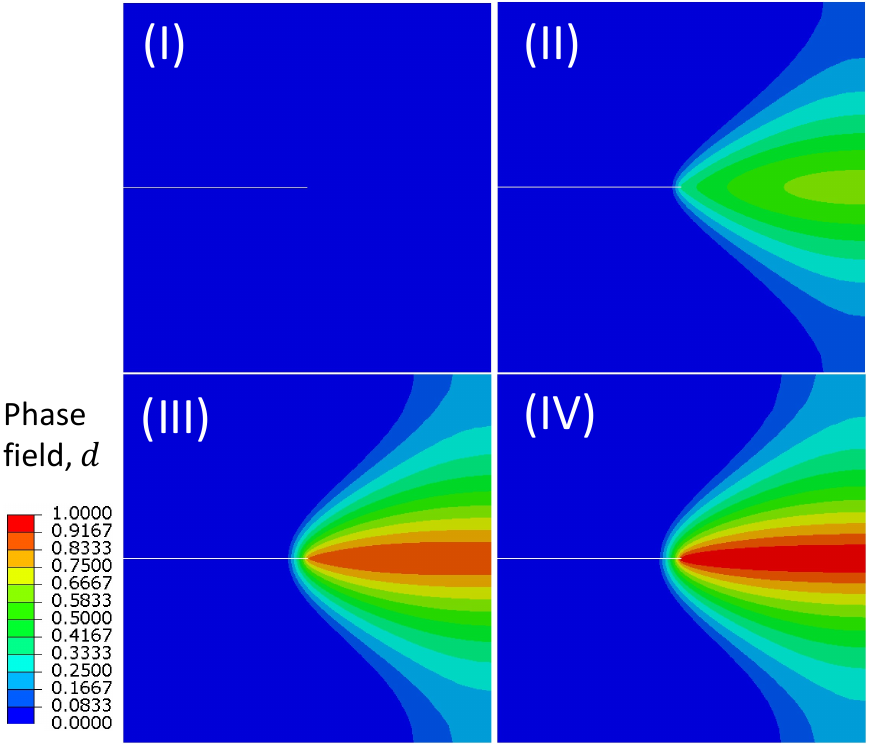} 
            \label{fig: SN_UT_60_Ctr}
        }%
        \hspace{0.2cm}
        \subfloat[]{%
            \includegraphics[scale=0.55]{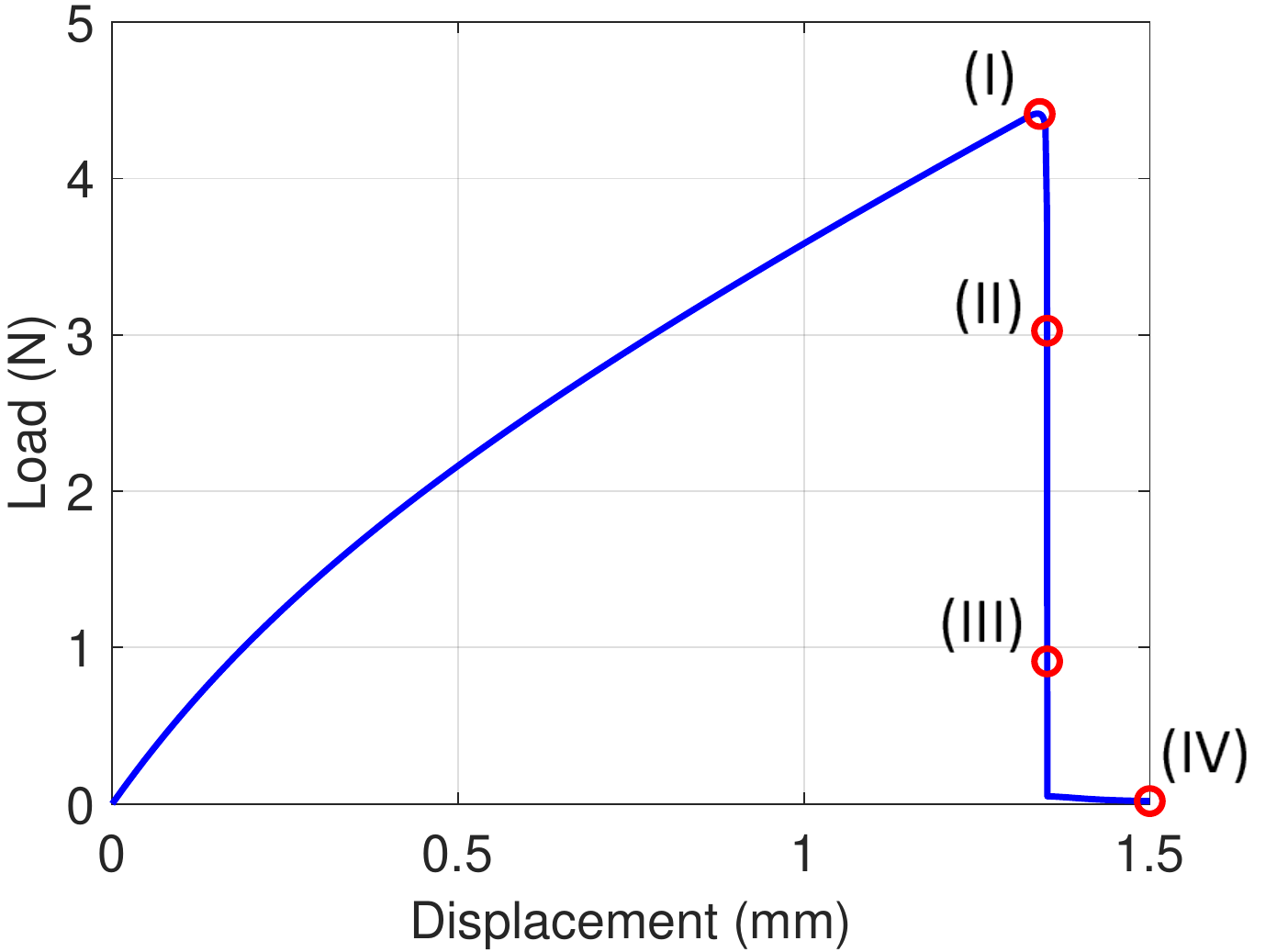} 
            \label{fig: SN_UT_60_LD}
        }%
    \caption{a) Sample geometry and boundary conditions for the single notch sample loaded in tension by a prescribed displacement $u$ (dimension are given in $mm$ unit), b) Cauchy stress $T_{22}^s$ contour at the point of maximum load for the sample that has been aged for 60 days, c) corresponding phase-field contours at various points during the simulation, and d) load-displacement curve.}
    \label{fig: SingleNotch_UniaxialTension}
\end{figure*}

In this section, we discuss the effects of aged material properties on the response of specimens containing pre-existing cracks. In particular, we investigate the case of a thermo-chemically aged single-notched specimen loaded under uniaxial tension as shown in Figure~\ref{fig: BCSNUT}. We assume that the specimen underwent the exact same aging procedure reported in the work of \cite{rezig2020thermo} and therefore the evolution of the crosslink density yields the exact same material properties highlighted in Table~\ref{tb: material properties}. We ran three simulations corresponding to three different aging states: 45, 60, and 85 days. The 45-day and 60-day simulations serve to demonstrate the effect of the material properties determined in Section~\ref{sec: validation}. The 85-day simulation serves as a parametric case designed to highlight the predictive capability of the proposed constitutive framework. Note that the crosslink density associated with the 85-day case was obtained through the linear extrapolation discussed above (see Figure~\ref{fig: crosslink_density}); the corresponding material properties were therefore determined based on the evolution functions established earlier.  

Figure~\ref{fig: SN_UT_60_S22} shows the crack-tip stress field contour (i.e., $T_{22}^s$) at the point of maximum load for the sample that has been aged for 60 days. The constitutive framework captures the stress concentration at the crack-tip correctly as it is typically known for the stress to localize at points of discontinuity. Figure~\ref{fig: SN_UT_60_Ctr} shows the evolution of the corresponding phase-field contour at select points along the load-displacement curve (Figure~\ref{fig: SN_UT_60_LD}). The effect of the length-scale as a material parameter is clearly demonstrated through the width of the crack band as it evolves during the simulation. Due to the increased length-scale for the present case, the width of the smeared crack appears to be rather large compared to the specimen's dimensions. Therefore, with the version of the phase-field employed in this work, physical interpretation of the size of the crack band (or equivalently the damage pattern) is to be approached with care when the length-scale is large with respect to the specimen's dimensions. Again, this observation has been pointed out in the work of \cite{mandal2019length} who confirmed that when the present phase-field version is employed, damage patterns only provide meaningful insight when the length scale is small with respect to the specimen's dimensions. Nonetheless, the load-displacement curve (Figure~\ref{fig: SN_UT_60_LD} correctly highlights the sudden drop and brittle fracture response that is typically observed for the single-notch example when loaded under uniaxial tension.

\begin{figure*}[hbt!]
    \centering
    \includegraphics[scale=0.62]{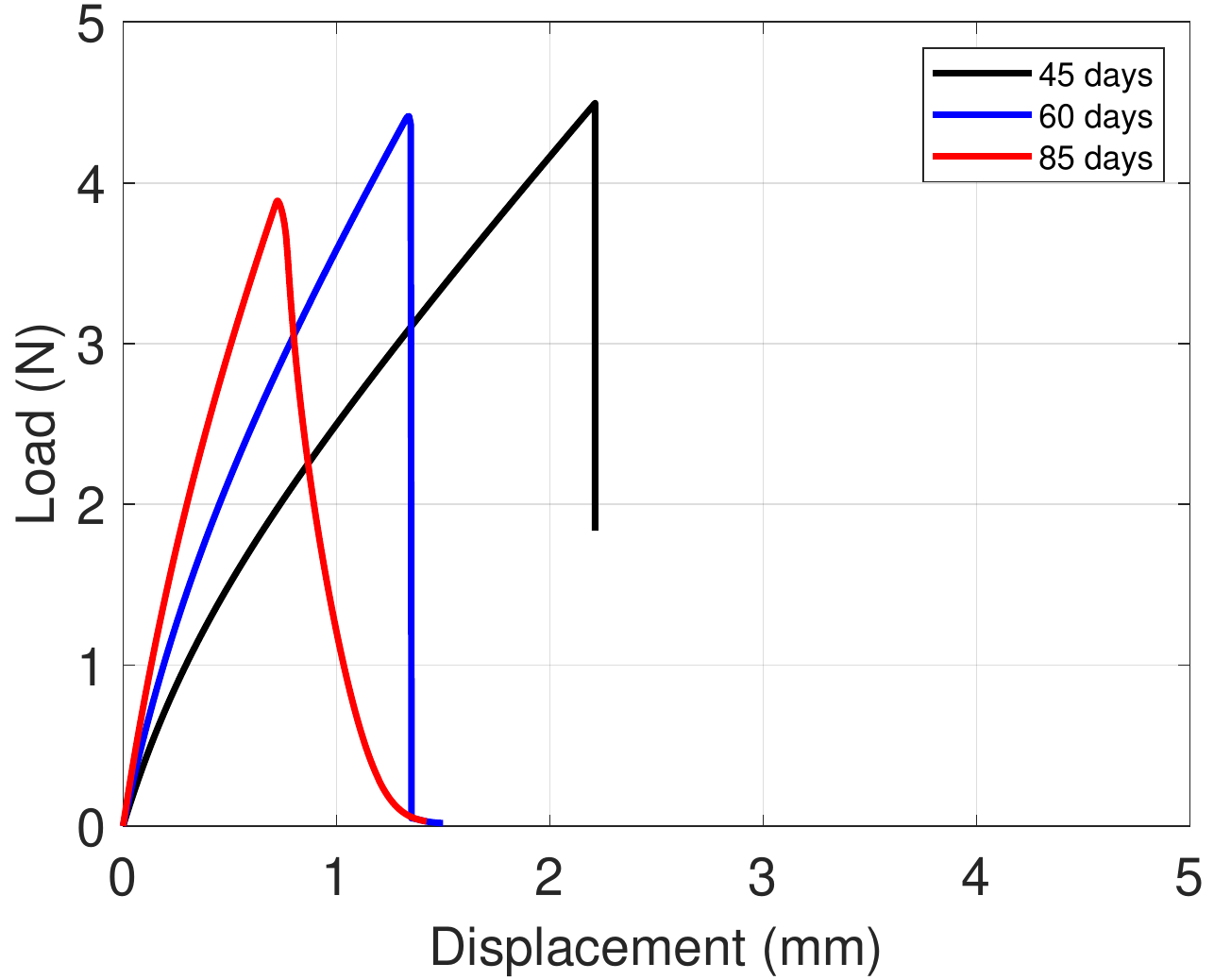} 
    \caption{Load-displacement curves corresponding to three aging states (i.e., 45, 60, and 85 days) for a single-notch sample loaded in uniaxial tension.}
    \label{fig: load_disp_SN_UT}
\end{figure*}

Figure~\ref{fig: load_disp_SN_UT} shows the load-displacement curves corresponding to the three aging states plotted together (i.e., 45, 60, and 85 days). At first glance, the figure demonstrates that the developed framework can predict the response of the 85-day aging time accurately as the associated load-displacement curve falls below that corresponding to the two other lower aging times as one would correctly predict.  
In a more in-depth analysis, it is clear that both the maximum load as well as the displacement at failure decrease with an increase in aging time. However, the stiffness increases with increasing aging time. The observed behavior for the three aging times is expected. Specifically, the decrease in maximum load and displacement at failure is governed by the evolution of the critical energy release rate which was shown to decrease with respect to aging time according to Eq.~(\ref{eq: fracture toughness 2}). The decrease in the critical energy release rate is itself due primarily to the fact that the number of Kuhn monomers decrease over aging time. Therefore, the premature fracture of thermo-chemically aged elastomers is directly linked to the decrease in the monomer density per chain. This observation implies that in the aged elastomer, a newly formed network containing shorter chains (albeit more rigid) compared to the original unaged network is continuously formed. On the other hand, the rise in stiffness is expected since the newly formed network contains a denser and a more crosslinked chain coil. In other words, while the chains in the aged elastomer are smaller and contribute to premature failure, the increase in crosslink density affect the stiffness and causes the material to undergo embrittlement. 

\section{Conclusions} \label{sec: Conclusions}

A physics and chemistry-based and thermodynamically consistent constitutive framework for the responses of thermo-chemically aged elastomers coupled with the phase-field approach to fracture has been proposed. The constitutive framework combines our understanding of how thermo-chemical aging affects the macromolecular structure of the rubber network and the adaptability of phase-field approach to simulate brittle fracture. The effect of thermally-driven crosslinking processes in modifying the bulk hyperelastic energy and the dissipated energy through fracture was considered. The framework was shown to be self-contained as it required the identification of only four material properties whose evolution during thermo-chemical aging was characterized entirely by the change in the crosslink density. Specifically, we showed that the evolution of the AB hyperelastic free energy (which is characterized by two micromechanically-motivated material properties: rubber modulus and number of Kuhn monomers), the critical energy release rate, the strain at fracture, and the length scale can be predicted entirely in terms of a single physio-chemical quantity: crosslink density. 

The interconnection between relevant material properties was discussed analytically for the case of 1D uniaxial tension. In particular, it was shown that the length-scale variable characterizing phase-field based damage models can be treated as an intrinsic constitutive material property and fracture nucleation in thermo-chemically aged elastomers can be captured conveniently through a strain-based criterion for crack initiation. The framework was subsequently implemented numerically through a user-element subroutine (UEL) on the FE software Abaqus to simulate more complicated geometries within a 2D context. The proposed framework was shown to predict the mechanical responses of thermo-chemically aged elastomers independently of any mechanical tests on the aged samples with very high accuracy. Such development is unprecedented in the literature particularly as the proposed framework is fairly simple and requires very few model parameters whose evolution during thermo-chemical aging can be connected directly to the evolving chain network characterized by the crosslink density.

A possible window for future development is to consider diffusion of chemical species such as oxygen inside the material and connect chemical gradients to the spatial variation of the material properties. Another window for improvement is to revisit fracture nucleation in such highly aged materials when loaded under complex deformation states especially for cases where test specimens are smooth (i.e., with no pre-existing cracks). Such important considerations are topics of on-going work. 

\begin{appendices}

\section{Phase-field approximation of crack discontinuities}
The phase-field approach was developed based on the diffuse representation of the localized discontinuity to simulate a fracture \citep{francfort1998revisiting,Ambati2015review,Miehe2010}. The principal idea governing the phase-field approach is that discontinuity is approximated by a smeared damage field. In this method, the crack surface is approximated with a variable, $d \in [0,1] \in \mathbb{R}$. If the phase-field is 0, the domain is intact or not-damaged, while if its value reaches 1, the crack is emerged, and the material has lost all of its resistance. 

A regularized crack surface functional, which measures a spatially regularized total crack surface, is defined in this approach as
\begin{equation} \label{Eq_fracture surface density}
\Gamma \left(d \right) = \int _{\Omega_0}  \gamma_{\left(d, \nabla d \right)} dd
\end{equation}
where $\Omega_0$ is the reference configuration of a material body, and $\gamma_{\left(d, \nabla d \right)}$ is the crack surface density functional expressed as:
\begin{equation} \label{Eq_crack surface density function}
\gamma_{\left(d, \nabla d \right)} = \frac{1}{c_{\alpha}} \left[ \frac{1}{l_c} \alpha_{(d)} + l_c \nabla d \cdot \nabla d \right] 
\quad\text{with}\quad
c_{\alpha} = 4 \int_0^1 \sqrt{\alpha_{(\beta)}} d\beta
\end{equation}
where $l_c$ is an incorporated length-scale for regularization and $\alpha_{(d)} \in [0,1]$ is a monotonically increasing crack geometric function satisfying the properties $\alpha(0)=0$ and $\alpha(1)=1$. The length-scale $l_c$ controls the diffuse damage field. In the limit of $l_c \to 0$, the original Griffith theory for fracture is recovered.

\section{Principle of virtual power for a solid medium with phase-field} \label{AppendixB virtual power}
According to the principle of virtual power, the external expenditure of virtual power should be balanced by the internal expenditure of virtual power
\begin{equation} \label{Eq_Principle_virtual_power}
\delta {\mathcal{P}_{int}}\left( {{\Omega _0}} \right) - \delta {\mathcal{P}_{ext}}\left( {{\Omega _0}} \right) =0
\end{equation}

In this study, the thermodynamic framework of \cite{Gurtin200347} along with the consideration of damage internal state variables are used to express the internal and external expenditures of power. Although the definition of the internal and external power is thought to be fixed, \citet{FREMOND19961083} and \citet{Fremond2002} demonstrated that it needs some modification to account for the effects of damage. A similar approach is used in this study to derive constitutive relationships that couple damage response of a solid. 

The internal expenditure of power which takes into account the microscopic
movements in the reference configuration, $\Omega_0$, of the solid can be characterized as \citep{FREMOND19961083}
\begin{equation} \label{Eq_internal_power}
{\mathcal{P}_{{\mathop{\rm int}} }} = \int\limits_{{\Omega _0}} {\left( {\mathbf{P} ^s} \cdot \dot {\mathbf{F}^s} + B \dot d + \mathbf{H} \cdot \nabla \dot d \right) d{\Omega _0}} 
\end{equation}
where $\mathbf{P}^s$ is the first Piola-Kirchoff stress tensor in the solid phase, $\mathbf{F}^s$ is the deformation gradient tensor of solid. The two non-classical quantities; $B$, is the internal work of damage (dual to $d$) and $\mathbf{H}$, is the flux vector of internal work of damage (dual to $\nabla d$).

\textbf{Note}. A generalized thermodynamic force conjugate to temperature can be included within internal power. However, this generalized thermodynamic force must be zero unless there exists a mechanism absorbing energy like micro-damage healing. Since micro-damage healing is not considered in the current study, the authors did not include this generalized thermodynamic force in the definition of internal power. It should be emphasized that the heat terms are incorporated in the formulation through the definition of the Clausius-Duhem inequality presented in the next sub-section.

External expenditure of power is defined in terms of the macroscopic body force vector, $\mathbf{f}_0$, macroscopic surface traction on solid skeleton, $\mathbf{t}_0^s$ as follows \citep{FREMOND19961083}
\begin{equation} \label{Eq_external_power}
\begin{array}{l}
{\mathcal{P}_{ext}} = {\mathcal{P}_{def}} = \int\limits_{{\Omega_0}} { {{\rho_0} \mathbf{f}_0 \cdot {\mathbf{v}_0^s} }  d{\Omega_0}} 
+ \int\limits_{{\Gamma_0}} { {\mathbf{t}_0^s} \cdot {\mathbf{v}_0^s} d{\Gamma_0}} + \int\limits_{{\Omega_0}} { {{\rho_0} A  \dot d }  d{\Omega_0}} 
+ \int\limits_{{\Gamma_0}} { b \dot d d{\Gamma_0}} \\
- \int\limits_{{\Omega_0}} { {{\rho_0} {\pmb{\gamma}_0^s} \cdot {\mathbf{v}_0^s} } d{\Omega_0}}  - \int\limits_{{\Omega_0}} { {\rho_0 \ddot d \cdot \dot d} \: d{\Omega_0}}
\end{array}
\end{equation}
where ${\rho_0}$ is the solid phase density, ${\mathbf{v}_0^s}$ is the velocity vectors of the solid, and ${\pmb{\gamma}_0^s}$ is the acceleration vectors of the solid. $A$ and $b$ are respectively the volumetric and surface external sources of damage work. A source of damage work $A$ or $b$ can be produced by chemical (or in some cases electrical) actions which break the links inside a material without macroscopic deformations. The quantity $\rho_0 \ddot d$ stands for the acceleration forces of the microscopic links. It should be mentioned that in this equation, the higher-order micro-traction at the evolving boundaries of the damaged regions are neglected. Moreover, $A$, and $b$ are considered to be zero here. First, this appendix section develops the framework for a solid medium with phase-field damage due to mechanical load, and diffusion terms and energies are not incorporated. Second, in this work, we focus on the cases of thermo-chemically aged elastomers, where diffusion and degradation have taken place in a different timescale, and their effects are implicitly taken into account according to Section~\ref{sec: cont model development}.

Then, the virtual expenditure of internal power, $\delta \mathcal{P}_{{\mathop{\rm int}} }$, and external power, $\delta \mathcal{P}_{{\mathop{\rm ext}} }$, can be defined when virtual prescribed fields replace the prescribed fields. According to the principle of virtual power, the external expenditure of virtual power should be balanced by the internal expenditure of virtual power ($\delta \mathcal{P}_{int} = \delta \mathcal {P}_{ext}$). Substituting the prescribed fields with virtual prescribed fields, equalizing the internal and external virtual power, using the divergence theorem and knowing that the virtual prescribed fields may be arbitrary, the balance equations are obtained as
\begin{equation} \label{Eq_solid_balance ref config}
\Div \left( {\mathbf{P} ^s} \right) + {\rho_0} \left( {\mathbf{f}_0 - {\pmb{\gamma}_0 ^s}} \right)  = 0\,\,{\rm{in}}\,\,\Omega_0 \qquad {\rm{and}} \qquad {\mathbf{t}_0^s} = {\mathbf{P} ^s}\mathbf{m}\,\,{\rm{on}}\,\,\Gamma_0
\end{equation}
\begin{equation} \label{Eq_damage_balance ref config}
\Div \left( \mathbf{H} \right) -B = \rho_0 \ddot d \,\,{\rm{in}}\,\,\Omega_0 \qquad {\rm{and}} \qquad 0 = \mathbf{H}  \cdot \mathbf{m}\,\,{\rm{on}}\,\,\Gamma_0
\end{equation} 
where $\mathbf{m}$ is the unit normal vector to the reference configuration. Eqs.~(\ref{Eq_solid_balance ref config}) expresses the local stress equilibrium equation or the macro-force balance for solid and the macroscopic boundary traction for solid as the density of the surface forces introduced. Eq.~(\ref{Eq_damage_balance ref config}) defines mechanical damage micro-force balance, which was first introduced by \cite{FREMOND19961083}, is used in this work to derive the phase-field nucleation and growth conditions. In the numerical simulations presented later, we neglect all inertial effects and body forces.

\section{Thermodynamic laws for a solid medium with phase-field} \label{AppendixC thermodynamic}

The first law of thermodynamics for a solid medium in the Lagrangian configuration considering the phase-field damage  is
\begin{equation} \label{Eq_First_thermo_final_Lag_localize_phaseField}
\rho_0 \dot E = \frac{1}{2}\mathbf{S} \cdot \dot{\mathbf{C}}  + B \dot d + \mathbf{H} \cdot \nabla \dot d + {\rho _0}R - \Div \left( {{\mathbf{Q}}} \right)
\end{equation}
where $E$ is Lagrangian specific internal energy, $\mathbf{S}$ is the second Piola-Kirchhoff stress tensor, ${\mathbf{C}}$ is the Right Cauchy-Green strain tensor, ${\mathbf{Q}}$ is Lagrangian heat flux, and $R$ is the specific Lagrangian heat production of the media.

The entropy inequality or the second law of thermodynamic is
\begin{equation} \label{eq: second law}
\rho_0 \dot Z \geq \frac{\rho R_0}{T}  - \Div (\frac{\mathbf{Q}}{T})
\end{equation}
where $Z$ and $T$ are the specific entropy and temperature of the media.

Substituting the first law into the second law, the Clausius-Duhem inequality can be obtained as
\begin{equation} \label{eq: CD}
\rho_0 \dot Z T + \frac{1}{2}\mathbf{S} \cdot \dot{ {\mathbf{C}}} + B \dot d + \mathbf{H} \cdot \nabla \dot d - \rho_0 \dot E  - \frac{\mathbf{Q} \cdot \nabla T}{T} \geq  0  
\end{equation}
Knowing that the specific Helmholtz free energy is $\Psi = E - TZ$, and substituting it into Eq.~(\ref{eq: CD}) gives
\begin{equation} \label{eq:CD_PF}
\frac{1}{2} \mathbf{S} \cdot \dot{{\mathbf{C}}} + B \dot d + \mathbf{H} \cdot \nabla \dot d - \rho_0 \dot \Psi - \rho_0 \dot T Z  -  \frac{\mathbf{Q} \cdot \nabla T}{T} \geq 0
\end{equation}

In developing the thermodynamic-based constitutive relationships, it is assumed that the state of material is characterized by suitable internal state variables. These variables implicitly describe important microstructural mechanisms that affect the macroscopic behavior of the material under specific loading and (initial) boundary conditions. The Helmholtz free energy is considered as the thermodynamic state potential depending on the internal state variables. In this study, we assume
\begin{equation} \label{eq:Helmholtz function_PF}
\Psi = \Psi \left( { {\mathbf{C}}}, d, \nabla d, T \right)
\end{equation}

Using the chain rule to take derivative of the Helmholtz free energy and substitute it in Eq. (\ref{eq:CD_PF}) provides
\begin{equation} \label{eq: Helmholtz in CD}
\left(\frac{1}{2} \mathbf{S} - \rho_0 \frac{\partial \Psi}{\partial {\mathbf{C}} }  \right)  \cdot \dot{{\mathbf{C}}}  + \left( B - \rho_0 \frac{\partial \Psi}{\partial d}  \right) \dot d + \left( \mathbf{H}  - \rho_o \frac{\partial \Psi}{\partial \nabla d}  \right)\cdot  {\nabla \dot d}  -  \rho_0 \left(  \frac{\partial \Psi}{\partial T} + Z \right) \dot T   -  \frac{\mathbf{Q} \cdot \nabla T}{T} \geq 0
\end{equation}

All processes and physical reactions should satisfy the Clausius–Duhem inequality as an accepted thermodynamic requirement. Thereafter, the internal energy should be conjectured properly to comply with the energy dissipation inequality and obtain the constitutive laws. \citet{zieglerIntroduction} stated that the correct estimation of energy dissipation requires decomposition of conjugate forces into energetic (or quasi-conservative, or non-dissipative) and dissipative components. 
Ziegler’s decomposition of conjugate forces into the energetic and dissipative components is used in this study to obtain the constitutive relationships. Heat transfer equation  can also be obtained by assuming proper forms for Helmholtz free energy and the rate of energy dissipation without decomposing the conjugate forces since their governing equations depend on advection and potential gradient and not time.

In order to obtain non-zero dissipation resulting from the solid dissipative processes, the following energetic and dissipative thermodynamic conjugate forces are defined from Eq.~(\ref{eq: Helmholtz in CD})
\begin{equation} \label{eq: sigma_energetic}
\mathbf{S}^{ene} = 2 \rho_0 \frac{\partial \Psi}{\partial {\mathbf{C}} }, \quad \text{and} \quad \mathbf{S}^{dis} = \mathbf{S}  - \mathbf{S}^{ene} 
\end{equation}
\begin{equation} \label{eq:B-energetic}
B^{ene} = \rho_0 \frac{\partial \Psi}{\partial d }, \quad \text{and} \quad B^{dis} = B  - B^{ene} 
\end{equation}
\begin{equation} \label{eq:H-energetic}
\mathbf{H}^{ene} = \rho_0 \frac{\partial \Psi}{\partial \nabla d }, \quad \text{and} \quad \mathbf{H}^{dis} = \mathbf{H}  - \mathbf{H}^{ene} 
\end{equation}
\begin{equation} \label{eq: entropy}
Z= -\frac{\partial \Psi}{\partial T} 
\end{equation}

Assuming that the solid state of material is hyperelastic and the Helmholtz free energy take the form of \citep{MIEHE201493}
\begin{equation}  \label{eq:Helmholtz}
\Psi \left( \mathbf{C}, d\right) = \omega \left(d\right) \Psi \left( \mathbf{C}\right)
\end{equation}
there will be no dissipation due to hyperelastic deformation (i.e., $\mathbf{S}^{dis} =0$), and the total dissipated energy become
\begin{equation} \label{eq: Helmholtz in CD_disConjForces}
\Pi = B^{dis}  \dot{d}  +\mathbf{H}^{dis} \cdot  \nabla \dot{d} -  \frac{\mathbf{Q} \cdot \nabla T}{T} \geq 0
\end{equation}

Here, we will use \textbf{the maximization of rate of energy dissipation criterion}, which states that over all possible different material responses, the naturally happening one is the one that maximizes the energy dissipation rate. \citet{zieglerIntroduction} elaborated on the validity and range of applicability of the maximum rate of energy dissipation in describing the natural behavior of materials. Although the maximum rate of energy dissipation is not a fundamental principle--and there are other methodologies depending upon the process--it has been used extensively in the literature to explicate various types of material behavior. In this study, the rate of energy dissipation maximization criterion is used. Thus, first, the energy dissipation in Eq.~(\ref{eq: Helmholtz in CD_disConjForces}) can be decomposed into the dissipation due to damage and thermal.
\begin{equation} \label{eq:Dissipation_decomposition_elasticDam}
\Pi = \Pi^{d} + \Pi^{th}  \geq 0
\end{equation}
Now, the constraint conditions should be applied to maximize the components of rate of energy dissipation functions such that
\begin{gather} \label{eq:Dissipation_constraints_elasticDam2}
D^d =   \Pi^{d} - \left( B^{dis}  \dot{d} +\mathbf{H}^{dis} \cdot \nabla \dot{d} \right) = 0 \\
D^{th} =   \Pi^{th} +  \frac{\mathbf{Q} \cdot \nabla T}{T}  = 0 
\end{gather}

We use Lagrange multiplier method to the objective function $\Omega^i = \Pi^i - l^i D^i$ ($i=d$, and $th$), where $l^i$ are associated Lagrange multipliers. Then, applying the necessary condition for maximizing the objective functions yield

\begin{equation}  \label{eq:Dissipation_Lagrange_Dam}
B^{dis} = \left(1-\frac{1}{\lambda^d} \right)\frac{\partial \Pi^d}{\partial \dot d}   \\
\end{equation}
\begin{equation}  \label{eq:Dissipation_Lagrange_gradDam}
\mathbf{H}^{dis} = \left(1-\frac{1}{\lambda^d} \right)\frac{\partial \Pi^{d}}{\partial \nabla \dot d} 
\end{equation}
\begin{equation}  \label{eq:Dissipation_Lagrange_Temp}
\nabla T = - \left(1-\frac{1}{\lambda^{th}} \right)\frac{\partial \Pi^{th}}{\partial \frac{\mathbf{Q} }{T}}   
\end{equation}

Substituting Eq.~(\ref{eq:Dissipation_Lagrange_Dam}) into $\Pi^d=B^{dis} \dot d + \mathbf{H}^{dis} \cdot \nabla \dot d$ gives $\lambda^d$ which is a constant, and similarly for the thermal Lagrange multiplier.  

\section{Constitutive equations of a solid medium with phase-field}
The presented thermodynamic framework is used at this point to derive the constitutive equations governing the responses of a hyperelastic material under mechanical damage. To obtain the constitutive equations, accurate Helmholtz free energy and rate of energy dissipation functions need to be speculated. The AB hyperelastic Helmholtz free energy \citep{ArrudaBoyce93} and the phase-field degradation function \citep{MIEHE201493} are given by
\begin{equation} \label{eq:HelmholtzArruda}
\begin{aligned} 
\Psi \left(\mathbf{C}\right) = \mu_0 N_0\Bigg[ \frac{\lambda_{chain}(\mathbf{C})}{\sqrt{N_0}} \mathcal{L}^{-1}\Big(\frac{\lambda_{chain}(\mathbf{C})}{\sqrt{N_0}}\Big) + ln\frac{\mathcal{L}^{-1}\Big(\frac{\lambda_{chain}(\mathbf{C})}{\sqrt{N_0}}\Big)}{\rm{sinh}(\mathcal{L}^{-1}\Big(\frac{\lambda_{chain}(\mathbf{C})}{\sqrt{N_0}}\Big))} \Bigg]
\end{aligned}
\end{equation}
\begin{equation}  \label{eq:Omega_d}
\omega \left(d\right) = (1-d)^2
\end{equation}

The next step is to assume the form of energy dissipation. Based on the phase-field theory, the rate of energy dissipation is equal to the critical energy release rate, $G_c$, times the rate of the crack surface density function \citep{Miehe2010}. Therefore, we write 
\begin{equation}  \label{eq:Dissipation}
\Pi \left(d, \nabla d \right) =  G_c \dot \gamma \left(d, \nabla d \right) 
\end{equation}
Taking the derivative of Eq.~(\ref{Eq_crack surface density function}) and substituting it into Eq.~(\ref{eq:Dissipation}) gives
\begin{equation}  \label{eq:DissipationPhaseField}
\Pi \left(d, \nabla d \right) =  G_c \frac{1}{c_{\alpha}} \left[ \frac{1}{l_c} {\alpha \prime}_{(d)}  \dot d + 2 l_c \nabla d \cdot \nabla \dot d \right]
\end{equation}
Substituting Eqs.~(\ref{eq:HelmholtzArruda}) and (\ref{eq:Omega_d}) into (\ref{eq: sigma_energetic}) gives the Second Piola-Kirchhoff stress tensor based on the AB model and the damage function $\omega \left( d \right)$. In addition, substituting Eqs.~(\ref{eq:HelmholtzArruda})-(\ref{eq:Omega_d}) and (\ref{eq:DissipationPhaseField}) into Eqs.~(\ref{eq:B-energetic})-(\ref{eq:H-energetic}) and (\ref{eq:Dissipation_Lagrange_Dam})-(\ref{eq:Dissipation_Lagrange_gradDam}) generates the energetic and dissipative conjugate forces of $B$ and $\mathbf{H}$ as
\begin{align} \label{eq:B-energetic_sub}
B^{ene} = \rho_0 \frac{\partial \Psi}{\partial d }= \rho_0 \omega \prime _{(d)} \frac{\partial \Psi}{\partial \omega }
\end{align}
\begin{equation}  \label{eq:Dissipation_Lagrange_Dam_sub}
B^{dis} = \left(1-\frac{1}{\lambda^d} \right)\frac{\partial \Pi^d}{\partial \dot d}= \left(1-\frac{1}{\lambda^d} \right) \frac{G_c {\alpha \prime}_{(d)}}{l_c c_{\alpha}}
\end{equation}
\begin{align} \label{eq:H-energetic-sub}
\mathbf{H}^{ene} = \rho_0 \frac{\partial \Psi}{\partial \nabla d }=0
\end{align}
\begin{equation} \label{eq:H-dis-sub}
\mathbf{H}^{dis} = \left(1-\frac{1}{\lambda^d} \right)\frac{\partial \Pi^{\nabla d}}{\partial \nabla \dot d} =  \left(1-\frac{1}{\lambda^d} \right) \frac{2l_c G_c}{c_{\alpha}} \nabla d   
\end{equation}
Therefore, according to the second part of Eqs.~(\ref{eq:B-energetic})-(\ref{eq:H-energetic}) 
\begin{equation}  \label{eq:Btotal}
B = \rho_0 \omega \prime _{(d)} \frac{\partial \Psi}{\partial \omega } + \frac{G_c {\alpha \prime}_{(d)}}{l_c c_{\alpha}}
\end{equation}
\begin{equation}  \label{eq:Htotal}
\mathbf{H} = \frac{2l_c G_c}{c_{\alpha}} \nabla d
\end{equation}
Notice that since $\lambda^d$ is a constant, it can be integrated into the other constants. Now, substituting Eqs.~(\ref{eq:Btotal}) and (\ref{eq:Htotal}) into the damage balance equation (i.e., Eq.~(\ref{Eq_damage_balance ref config})) gives
\begin{align} \label{Eq_damage_balances}
\frac{2l_c G_c}{c_{\alpha}} \Delta d - \rho \omega \prime _{(d)} \frac{\partial \Psi}{\partial \omega } -  \frac{G_c {\alpha \prime}_{(d)}}{l_c c_{\alpha}}  = \rho \ddot d \,\,{\rm{in}}\,\,\mathcal{B} \quad {\rm{and}} \quad \mathbf{0}= \frac{2l_c G_c}{c_{\alpha}} \nabla d  \cdot \mathbf{n}\,\,{\rm{on}}\,\,{\Gamma_0}_{dt}
\end{align}
which is similar to the damage balance equation used in other works such as \citep{wu2017unified,MANDAL2020107196,wu2020variationally}. It has been shown here that the phase-field equations can be obtained systematically within the \cite{FREMOND19961083} framework. Moreover, assuming the form of thermal dissipative energy to be $\Pi^{th}=K \mathbf{Q} \cdot \mathbf{Q}$, where $K$ is the thermal conductivity, generates Fourier heat conduction law. It also must be emphasized again that in this work, we neglected the energies due to oxygen diffusion because of the difference in the mass diffusion time frame and the mechanical damage response. However, the equations can be easily modified to include the stored and dissipative energies due to mass diffusion. For a detailed derivation of such problem please refer to the previous work of the author \citep{SHAKIBA201653}. 


\section{Finite element implementation} \label{Apdx: FE implementation}

In this section, the finite element (FE) implementation of the proposed phase-field model approach is described. We begin by establishing the weak-forms associated with coupled displacment-damage problem. Then the finite-element discretization and piecewise approximation corresponding to the displacement and damage fields are established and the resulting discrete equations are provided.

\subsection{Weak forms}

In accordance with standard practice, by considering an arbitrary vector field $\mathbf{w_u}$ whose components vanish at the corresponding essential boundary segments, the weak form corresponding to the displacement field can be written as follows:

\begin{equation} \label{weak_form_displacement}
\begin{gathered}
    \int_{{\Omega_u}_0} \mathbf{w_u} \cdot \rho (\mathbf{f}_0 - {\pmb{\gamma}_0 ^s}) dV + \int_{{\Omega_u}_0} \bar{\mathbf{F}}_u : \mathbf{P}^s dV = \int_{{\Gamma_d}_0}  \mathbf{w_u} \cdot \mathbf{t}_0^s dA \\ 
    \forall \mathbf{w_u} \ \  \rm{with} \ \ \mathbf{w_u} = \mathbf{0} \ \rm{on} \ {{\Gamma_u}_0}
\end{gathered}
\end{equation} 
where we define $\bar{\mathbf{F}}_u = \frac{\partial \mathbf{w_u}}{\partial \mathbf{X}} $, i.e., the partial derivative of the arbitrary vector field $\mathbf{w_u}$ with respect to the reference coordinates $X$.

Equivalently, by considering an arbitrary vector field $\mathbf{w_d}$ whose components vanish at the corresponding essential boundary segments, the weak form corresponding to the damage field can be expressed as follows:

\begin{equation}\label{weak_form_damage}
\begin{gathered} 
    \int_{{\Omega_d}_0} \mathbf{w_d} \cdot \mathbf{B} dV + \int_{{\Omega_d}_0} \bar{\mathbf{F}}_d \cdot \mathbf{H} dV + \int_{{\Omega_d}_0} \mathbf{w_d} \cdot \rho \ddot{d} dV  = 0 \\
    \rm{or \ after \ substitution} \\
    \int_{{\Omega_d}_0} \mathbf{w_d} \cdot (\rho \omega \prime _{(d)} \frac{\partial \Psi}{\partial \omega } + \frac{G_c {\alpha \prime}_{(d)}}{l_c c_{\alpha}} ) dV + \int_{{\Omega_d}_0} \bar{\mathbf{F}}_d \cdot  (\frac{2l_c G_c}{c_{\alpha}} \nabla d ) dV + \int_{{\Omega_d}_0} \mathbf{w_d} \cdot \rho \ddot{d} dV  = 0 \\
    \forall \mathbf{w_d} \ \  \rm{with} \ \ \mathbf{w_d} = \mathbf{0} \ \rm{on} \ {{\Gamma_d}_0}
\end{gathered}    
\end{equation}
where similarly $\bar{\mathbf{F}}_d = \frac{\partial \mathbf{w_d}}{\partial \mathbf{X}} $, i.e., the partial derivative of the arbitrary vector field $\mathbf{w_d}$ with respect to the reference coordinates $X$.

\subsection{Finite element discretization}

The weak forms~\ref{weak_form_displacement} and ~\ref{weak_form_damage} are usually discretized using multi-field finite elements. Without loss of generality, in what follows, we consider the case of 2D problems, with the assumption that the 3D formulation extends in a straightforward manner. For such problems, the 2D domain $\Omega_o$ is discretized into subdomains called elements and each element consists of nodes. For 2D problems, the standard and most commonly employed element shapes, i.e., the three-node triangular element and the four-node quadrilateral element are used. For the coupled displacement-phase-field problem in hand, each element node has three nodal degrees of freedom: two for the displacement field and one for the damage field.

The displacement field $\textbf{\textit{u}(X)}$ and damage field $d(\textbf{X})$ are approximated using the nodal displacement and damage vectors, $u_i$ and $d_j$ through their corresponding shape functions, where $i \in [1,$ndof$_u]$ and $j \in [1,$ndof$_u]$ wherein ndof$_u$ and ndof$_d$ denote the number of degrees of freedom associated with the displacement and damage fields, respectively. The shape functions are established as functions of the Lagrangian coordinates \{$\mathbf{X_o}$\}. As such, the displacement and damage fields can be written as follows
\begin{align} \label{nodal_vectors}
\mathbf{u(X)} =  \sum_k^{nnode} \mathbf{N}_k\mathbf{u_k} \qquad {\rm{and}} \qquad \mathbf{d(X)} =  \sum_k^{nnode}  \mathbf{N}_k\mathbf{d_k} 
\end{align} 
where $nnode$ is the number of nodes in the element. Similarly, we use the same shape functions to approximate the arbitrary vector fields $\mathbf{w_u}$ and $\mathbf{w_d}$ and write:
\begin{align} \label{weight_nodal_vectors}
\mathbf{w_u(X)} =  \sum_k^{nnode} \mathbf{N}_k\mathbf{{w_u}_k} \qquad {\rm{and}} \qquad \mathbf{w_d(X)} =  \sum_k^{nnode}  \mathbf{N}_k\mathbf{{w_d}_k} 
\end{align} 

Substituting Eqs.(~\ref{nodal_vectors}) and (\ref{weight_nodal_vectors}) in Eqs.(~\ref{weak_form_displacement}) and (\ref{weak_form_damage}) yield the following element-level system of equations:

\begin{equation}
    \begin{gathered}
    \int_{{\Omega_0}^e} \bigg(\mathbf{P^s} \frac{\partial N_k}{\partial \mathbf{X}} \bigg) dV = \int_{{{\Gamma_u}_0}^e} N_k \mathbf{t_0^s} dA \\
    \int_{{\Omega_0}^e} \bigg(\mathbf{H} \cdot \frac{\partial N_k}{\partial \mathbf{X}} \bigg) dV + \int_{{\Omega_0}^e} \bigg(N_k \mathbf{B}  \bigg) dV = 0
    \end{gathered}
\end{equation}
where $\mathbf{H} = \frac{2l_c G_c}{c_{\alpha}} \nabla d $ and $\mathbf{B} = \rho \omega \prime _{(d)} \frac{\partial \Psi}{\partial \omega } + \frac{G_c {\alpha \prime}_{(d)}}{l_c c_{\alpha}}  $.

This system of coupled equations is solved iteratively using an appropriate numerical procedure by the defining the following element-level residuals for the displacement and phase-field

\begin{equation}
    \begin{gathered}
    \mathbf{R^e_u} = - \int_{{\Omega_0}^e} \bigg(\mathbf{P^s} \frac{\partial N_k}{\partial \mathbf{X}} \bigg) dV + \int_{{{\Gamma_u}_0}^e} N_k \mathbf{t_0^s} dA \\
    \mathbf{R^e_d} = \int_{{\Omega_0}^e} \bigg(\mathbf{H} \cdot \frac{\partial N_k}{\partial \mathbf{X}} \bigg) dV + \int_{{\Omega_0}^e} \bigg(N_k \mathbf{B}  \bigg) dV 
    \end{gathered}
\end{equation}
and using the corresponding tangents

\begin{equation}
    \begin{gathered}
    \mathbf{K^e_{uu}} = -\frac{\partial \mathbf{R^e_u}}{\partial \mathbf{u}} \ , \ \mathbf{K^e_{dd}} = -\frac{\partial \mathbf{R^e_d}}{\partial \mathbf{d}} 
    \end{gathered}
\end{equation}

In this work, the staggered scheme of \cite{Miehe2010} is used and the following system is solved iteratively using a Newton Raphson algorithm 

\begin{equation}
    \begin{gathered}
        \begin{bmatrix}
            \mathbf{K^e_{uu}} & 0 \\
            0 & \mathbf{K^e_{dd}} \\
        \end{bmatrix}_n
        \begin{bmatrix}
           \mathbf{u} \\
                    d \\
        \end{bmatrix}_{n+1} = -
        \begin{bmatrix}
           \mathbf{R^e_u} \\
            \mathbf{R^e_d} \\
        \end{bmatrix}_n
    \end{gathered}
\end{equation}
where the subscript $n$ refers to the converged step and $n+1$ denotes the next unknown step. 

\end{appendices}

\section*{Acknowledgement}
The authors gratefully acknowledge the support from the National Science Foundation under the award number CMMI-1914565, and the Air-Force Office of Scientific Research (AFOSR) Young Investigator Program (YIP) award \#FA9550-20-1-0281. The authors also acknowledge Advanced Research Computing at Virginia Tech for providing computational resources and technical support that have contributed to the results reported within this paper. URL: https://arc.vt.edu/


\bibliography{output}



\end{document}